\documentclass[journal]{IEEEtran}

\usepackage{cite}
\usepackage{url}
\usepackage{graphicx} 
\usepackage{amsmath,amssymb} 
\usepackage{algorithm} 
\usepackage{algorithmic}
\usepackage[normalem]{ulem}
\usepackage{multirow}
\usepackage{float}
\usepackage{color}
\usepackage{overpic}
\usepackage{rotating}
\usepackage[table]{xcolor}
\usepackage[pagebackref=false,breaklinks=true,colorlinks, bookmarks=false]{hyperref}
\usepackage{multirow}
\usepackage{makecell}
\usepackage{pgfplots}
\definecolor{G}{RGB}{124, 252, 0} 
\definecolor{R}{RGB}{0, 0, 255} 
\definecolor{Y}{RGB}{0, 255,255} 

\newcommand{\figref}[1]{Fig.~\ref{#1}}
\newcommand{\secref}[1]{Sec. \ref{#1}}
\newcommand{\tabref}[1]{Tab.~\ref{#1}}

\def\ourmodel{Polyp-PVT}
\def\etal{\textit{et al.}}

\begin{document}

\title{Polyp-PVT: Polyp Segmentation with Pyramid Vision Transformers}

\author{
Bo~Dong, 
Wenhai~Wang,
Deng-Ping Fan,
Jinpeng Li, 
Huazhu Fu,
and Ling Shao

\thanks{B.~Dong and D.-P.~Fan are with the Nankai University, Tianjin, 300350, China.
* Corresponding author (dengpfan@gmail.com).
}
\thanks{W.~Wang is with Shanghai Artificial Intelligence Laboratory, Shanghai, 200232, China}
\thanks{J.~Li is with Computer Vision Lab, Inception Institute of Artificial Intelligence, Abu Dhabi 00000, UAE}
\thanks{H.~Fu is with Institute of High Performance Computing, Agency for Science, Technology and Research, Singapore 138632, Singapore}
\thanks{L.~Shao is with UCAS-Terminus AI Lab, Terminus Group, Chongqing 400042, China}
}

\maketitle

\begin{abstract}

Most polyp segmentation methods use CNNs as their backbone, leading to two key issues when exchanging information between the encoder and decoder: 1) taking into account the differences in contribution between different-level features and 2) designing an effective mechanism for fusing these features.
Unlike existing CNN-based methods, we adopt a transformer encoder, which learns more powerful and robust representations.
In addition, considering the image acquisition influence and elusive properties of polyps, we introduce three standard modules, including a cascaded fusion module (CFM), a camouflage identification module (CIM), and a similarity aggregation module (SAM).
Among these, the CFM is used to collect the semantic and location information of polyps from high-level features; the CIM is applied to capture polyp information disguised in low-level features, and 
the SAM extends the pixel features of the polyp area with high-level semantic position information to the entire polyp area,
thereby effectively fusing cross-level features.
The proposed model, named \ourmodel, effectively suppresses noises in the features and significantly improves their expressive capabilities.
Extensive experiments on five widely adopted datasets show that the proposed model is more robust to various challenging situations (\emph{e.g.}, appearance changes, small objects, rotation) than existing representative methods. 
The proposed model is available at \url{https://github.com/DengPingFan/Polyp-PVT}.
\end{abstract}


\section{Introduction}
\label{sec:introduction}
Colonoscopy is the gold standard for detecting colorectal lesions since it enables colorectal polyps to be identified and removed in time, thereby preventing further spread. 
As a fundamental task in medical image analysis, polyp segmentation (PS) aims to locate polyps accurately in the early stage, which is of great significance in the clinical prevention of rectal cancer. 
Traditional PS models mainly rely on low-level features, \emph{e.g.}, texture~\cite{fiori2014complete}, geometric features~\cite{mamonov2014automated}, simple linear iterative clustering superpixels~\cite{maghsoudi2017superpixel}.
However, these methods yield low-quality results and suffer from poor generalization ability.
With the development of deep learning, PS has achieved promising progress.
In particular, the U-shaped~\cite{ronneberger2015unet} has attracted significant attention due to its ability to adopt multi-level features for reconstructing high-resolution results. 
PraNet~\cite{fan2020pranet} employs a two-stage segmentation approach, adopting a parallel decoder to predict rough regions and an attention mechanism to restore a polyp's edges and internal structure for fine-grained segmentation.
ThresholdNet~\cite{guo2020learn} is a confidence-guided data enhancement method based on a hybrid manifold for solving the problems caused by limited annotated data and imbalanced data distributions.

\begin{figure}[t!]
	\centering
    \small
    \vspace{8pt}
	\begin{overpic}[width=.93\linewidth]{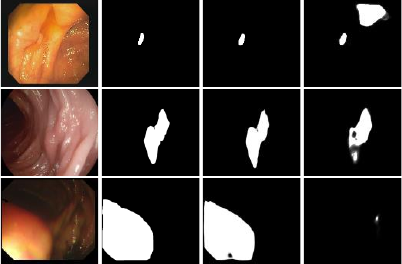}
	\put(7,-4.5){Image}
	\put(35,-4.5){GT}
	\put(58.5,-4.5){\color{red}Ours}
	\put(81.5,-4.5){SANet}
    \end{overpic}
    \vspace{5pt}
	\caption{
	The segmentation examples of our model and SANet~\cite{wei2021shallow} with different challenge cases, \emph{e.g.}, camouflage ($1^{st}$ and $2^{nd}$ rows) and image acquisition influence ($3^{rd}$ row). 
	The images from top to bottom are from ClinicDB~\cite{bernal2015wm}, ETIS~\cite{silva2014toward}, and ColonDB~\cite{tajbakhsh2015automated}, which show that our model has better generalization ability. }
    \label{figure:first_map}
\end{figure}
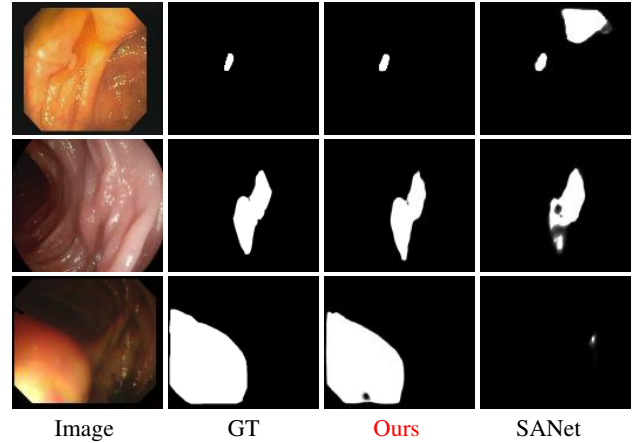

Although these methods have greatly improved accuracy and generalization ability compared to traditional methods, it is still challenging for them to locate the boundaries of polyps, as shown in \figref{figure:first_map}, for several reasons:
(1) \textcolor[RGB]{31,100,212}{Image noise.} During the data collection process, the lens rotates in the intestine to obtain polyp images from different angles, which also causes motion blur and reflector problems. As a result, this greatly increases the difficulty of polyp detection;
(2) \textcolor[RGB]{31,100,212}{Camouflage.} The color and texture of polyps are very similar to surrounding tissues, with low contrast, providing them with powerful camouflage properties~\cite{fan2021concealed,fan2020camouflaged}, and making them difficult to identify;
(3) \textcolor[RGB]{31,100,212}{Polycentric data.} Current models struggle to generalize to multicenter (or unseen) data with different domains/distributions.
To address the above issues, our contributions in this paper are as follows:

\begin{itemize}
\item We present a novel polyp segmentation framework, termed \ourmodel. Unlike existing CNN-based methods, we adopt the pyramid vision transformer as an encoder to extract more robust features.

\vspace{-5pt}
\item To support our framework, we introduce three simple modules. Specifically, the cascaded fusion module (CFM) collects polyps' semantic and location information from the high-level features through progressive integration. 
Meanwhile, the camouflage identification module (CIM) is applied to capture polyp cues disguised in low-level features, using an attention mechanism to pay more attention to potential polyps, reducing incorrect information in the lower features. 
We further introduce the similarity aggregation module (SAM) equipped with a non-local and convolutional graph layer to mine local pixels and global semantic cues from the polyp area.

\vspace{-5pt}
\item Finally, we conduct extensive experiments on five challenging benchmark datasets, including Kvasir-SEG~\cite{jha2020kvasir}, ClinicDB~\cite{bernal2015wm}, ColonDB~\cite{tajbakhsh2015automated}, Endoscene~\cite{vazquez2017benchmark}, and ETIS~\cite{silva2014toward}, to evaluate the performance of the proposed \ourmodel.
On ColonDB, our method achieves a mean Dice (mDic) of 0.808, which is 5.5\% higher than the existing cutting-edge method SANet~\cite{wei2021shallow}.
On the ETIS dataset, our model achieves a mean Dice (mDic) of 0.787, which is 3.7\% higher than SANet~\cite{wei2021shallow}.
\end{itemize}

\section{Related Works} 
\subsection{Polyp Segmentation}\label{subsec:polyp segmentation}
\textcolor[RGB]{31,100,212}{Traditional Methods.} {Computer-aided detection is an effective alternative to manual detection, and a detailed survey has been conducted on detecting ulcers, polyps, and tumors in wireless capsule endoscopy imaging~\cite{rahim2020survey}.}
Early solutions for polyp segmentation were mainly based on low-level features, such as texture~\cite{mamonov2014automated}, geometric features~\cite{mamonov2014automated}, or simple linear iterative clustering superpixels~\cite{maghsoudi2017superpixel}.
However, these methods have a high risk of missed or false detection due to the high similarity between polyps and surrounding tissues.

\textcolor[RGB]{31,100,212}{Deep Learning-Based Methods.} Deep learning techniques~\cite{he2016deep,simonyan2015very,li2019selective,wang2018mixed,long2015fully,cai2022using,tomar2022tganet,zhang2022lesion,shi2022polyp,zhao2022semi} have greatly promoted
the development of polyp segmentation tasks. 
Akbari \etal~\cite{akbari2018polyp} proposed a polyp segmentation model using a fully convolutional neural network, whose segmentation results are significantly better than traditional solutions.
Brandao \etal~\cite{brandao2018towards} used the shape from the shading strategy to restore depth, merging the result into an RGB model to provide richer feature representations.
More recently, encoder-decoder-based models, such as U-Net~\cite{ronneberger2015unet}, UNet++~\cite{zhou2018unet++}, and ResUNet++~\cite{JhaSRJLHJ19}, have gradually come to dominate the field with excellent performance.
Sun \etal~\cite{sun2019colorectal} introduced a dilated convolution to extract and aggregate high-level semantic features with resolution retention to improve the encoder network.
Psi-Net~\cite{murugesan2019psi} introduced a multi-task segmentation model that combines contour and distance map estimation to assist segmentation mask prediction.
Hemin \etal~\cite{qadir2019polyp} first attempted to use a deeper feature extractor to perform polyp segmentation based on Mask R-CNN~\cite{he2017mask}.

Different from the methods based on U-Net~\cite{ronneberger2015unet,zhou2018unet++,AlamTTJR20}, PraNet~\cite{fan2020pranet} uses reverse attention modules to mine boundary information with a global feature map, which is generated by a parallel partial decoder from high-level features.
Polyp-Net~\cite{banik2020polyp} proposed a dual-tree wavelet pooling CNN with a local gradient-weighted embedding level set, effectively avoiding erroneous information in high signal areas, thereby significantly reducing the false positive rate.
Rahim \etal~\cite{rahim2021deep} proposed to use different convolution kernels for the same hidden layer for deeper feature extraction with MISH and rectified linear unit activation functions for deep feature propagation and smooth non-monotonicity. In addition, they adopted joint generalized intersections, which overcome scale invariance, rotation, and shape differences.
Jha \etal~\cite{jha2021real} designed a real-time polyp segmentation method called ColonSNet.
For the first time, Ahmed \etal~\cite{ahmed2020generative} applied the generative adversarial network to the field of polyp segmentation.
Another interesting idea proposed by Thambawita~\etal~\cite{ThambawitaHHR20} is introducing pyramid-based augmentation into the polyp segmentation task.
Further, Tomar \etal~\cite{tomar2020ddanet} designed a dual decoder attention network based on ResUNet++ for polyp segmentation.
More recently, MSEG~\cite{huang2021hardnet} improved the PraNet and proposed a simple encoder-decoder structure. 
Specifically, they used Hardnet~\cite{chao2019hardnet} to replace the original backbone network Res2Net50 backbone network and removed the attention mechanism to achieve faster and more accurate polyp segmentation. 
As an early attempt, Transfuse~\cite{zhang2021transfuse} was the first to employ a two-branch architecture combining CNNs and transformers in a parallel style. 
DCRNet~\cite{yin2021duplex} uses external and internal context relations modules to separately estimate the similarity between each location and all other locations in the same and different images.
MSNet~\cite{Zhao_2021_MICCAI} introduced a multi-scale subtraction network to eliminate redundancy and complementary information between the multi-scale features.
Providing a comprehensive review of polyp segmentation is beyond the scope of this paper. 
In Tab.~\ref{tab: model_survey}, however, we briefly survey  representative works related to ours.

\begin{table*}[t!]
\small
  \centering
  \caption{
  A survey  on polyp segmentation. CL = CVC-CLINIC, EL = ETIS-Larib, C6 = CVC-612, AM = ASU-Mayo~\cite{zhou2017fine,tajbakhsh2016convolutional}, ES = EndoScene, DB = ColonDB, CV = CVC-VideoClinicDB, C = Colon, ED = Endotect 2020, KS = Kvasir-SEG, KCS = Kvasir Capsule-SEG, PraNet = same to datasets used in PraNet~\cite{fan2020pranet}, IS = image segmentation, VS = video segmentation, CF = classfication, OD = object detection, Own = private data. CSCPD~\cite{fiori2014complete}, APD~\cite{mamonov2014automated}, SBCP~\cite{maghsoudi2017superpixel}, FCN~\cite{akbari2018polyp}, D-FCN~\cite{brandao2018towards}, UNet++~\cite{zhou2018unet++}, Psi-Net~\cite{murugesan2019psi}, Mask R-CNN~\cite{qadir2019polyp}, UDC~\cite{sun2019colorectal}, ThresholdNet~\cite{guo2020learn}, MI2GAN~\cite{xie2020mi}, ACSNet~\cite{zhang2020ACSN}, PraNet~\cite{fan2020pranet}, GAN~\cite{ahmed2020generative}, APS~\cite{tomar2021automatic}, PFA~\cite{ThambawitaHHR20}, MMT~\cite{JhaHEJJLRH20}, U-Net-ResNet50~\cite{AlamTTJR20}, Survey~\cite{rahim2020survey}, Polyp-Net~\cite{banik2020polyp}, Deep CNN~\cite{rahim2021deep}, EU-Net~\cite{PatelBW21}, DSAS~\cite{lumini2021deep}, U-Net-MobileNetV2~\cite{branch2021polyp}, DCRNet~\cite{yin2021duplex}, MSEG~\cite{huang2021hardnet}, FSSNet~\cite{khadga2021few}, AG-CUResNeSt~\cite{sang2021ag}, MPAPS~\cite{yang2021mutual}, ResUNet++~\cite{JhaSJLJHR21}, NanoNet~\cite{jha2021nanonet}, ColonSegNet~\cite{jha2021real}, Segtran~\cite{segtran}, DDANet~\cite{tomar2020ddanet}, UACANet~\cite{kim2021uacanet}, DivergentNet~\cite{divergentNets}, DWHieraSeg~\cite{guo2021dynamic}, Transfuse~\cite{zhang2021transfuse}, SANet~\cite{wei2021shallow}, PNS-Net~\cite{ji2021pnsnet}.}
    \renewcommand{\arraystretch}{1.1}
    \setlength\tabcolsep{4.8pt}
    \begin{tabular}{r|r|c|c|c|c|l}
    \hline
    No. & Model & Publication & Code  & Type & Dataset & Core Components\\
    \hline   
    1&CSCPD & IJPRAI& N/A &   IS    & Own &  Adaptive-scale candidate \\

   2 &APD & TMI  & N/A &   IS    & Own &  Geometrical analysis, binary classifier\\
    
    3&SBCP & SPMB  & N/A &   IS    & Own &  Superpixel\\

    4&FCN & EMBC & N/A &   IS    & DB&  FCN and patch selection\\    
    
    5&D-FCN  & JMRR  & N/A    &  IS     & CL, EL, AM, and DB   & FCN and Shape-from-Shading (SfS)\\
    
   6 &UNet++ &DLMIA    & \href{https://github.com/MrGiovanni/UNetPlusPlus}{PyTorch} &   IS    & AM & Skip pathways and deep supervision\\  
    
    7&Psi-Net& EMBC   & \href{https://github.com/Bala93/Multi-task-deep-network}{PyTorch} &   IS    & Endovis &  Shape and boundary aware\\  
    
   8& Mask R-CNN & ISMICT  & N/A &   IS    & C6, EL, and DB &  Deep feature extractors \\ 
    
   9 &UDC & ICMLA    & N/A &   IS    & C6 and EL &  Dilation convolution \\
    
    10&ThresholdNet & TMI     & \href{https://github.com/Guo-Xiaoqing/ThresholdNet}{PyTorch} &     IS  & ES and WCE   & \makecell[l]{Learn to threshold \\ Confidence-guided manifold mixup} \\
    
    11&MI2GAN  & MICCAI   & N/A    & IS      &C6 and EL  & GAN based model\\
   12 &ACSNet   & MICCAI  & \href{https://github.com/ReaFly/ACSNet }{PyTorch}& IS       &ES and KS   &  Adaptive context selection\\
    13&PraNet & MICCAI  & \href{https://github.com/DengPingFan/PraNet}{PyTorch} &  IS     & PraNet & Parallel partial decoder attention\\
    14&GAN  & MediaEval   & N/A    & IS      &  KS &  Image-to-image translation\\
    15&APS   & MediaEval   & N/A    &      IS & KS &Variants of U-shaped structure\\
    16&PFA & MediaEval  & \href{https://github.com/vlbthambawita/pyra-pytorch}{PyTorch} &   IS    & KS &  Pyramid focus augmentation\\
    17&MMT  & MediaEval  & N/A    &   IS    & KS & Competition introduction \\
    
    18&U-Net-ResNet50 & MediaEval   & N/A    & IS       &  KS & Variants of U-shaped structure\\
    19&Survey & CMIG  & N/A  & CF     & Own & Classification \\
    
   20&Polyp-Net & TIM  & N/A & IS     & DB and CV &  Multimodel fusion network \\
    
    21&Deep CNN & BSPC   & N/A& OD     & EL &   Convolutional neural network \\
    
    22&EU-Net & CRV  & \href{https://github.com/rucv/Enhanced-U-Net}{PyTorch} & IS      & PraNet &  Semantic information enhancement \\
    23&DSAS & MIDL  & \href{https://github.com/LorisNanni/Deep-ensembles-based-on-Stochastic-Activation-Selection-for-Polyp-Segmentation}{Matlab} &  IS     &  KS & Stochastic activation selection\\
    24&U-Net-MobileNetV2 & arXiv  & N/A    &    IS   & KS & Variants of U-shaped structure \\
   25 &DCRNet & ISBI  & \href{https://github.com/PRIS-CV/DCRNet}{PyTorch}&    IS   &\makecell[c]{ES, KS, and\\PICCOLO}    &\makecell[l]{Within-image \\and cross-image contextual relations}   \\
    26&MSEG & arXiv   & \href{https://github.com/james128333/HarDNet-MSEG}{PyTorch} &IS       & PraNet & Hardnet and partial decoder\\
    27&FSSNet & arXiv   & N/A   &       IS & C6 and KS &Meta-learning\\
    28&AG-CUResNeSt& RIVF   & N/A    &    IS   & PraNet &  ResNeSt, attention gates\\
   29 &MPAPS & JBHI   & \href{https://github.com/CityU-AIM-Group/MPA-DA}{PyTorch}   &       IS &  DB, KS, and EL & Mutual-prototype adaptation network \\  
    30&ResUNet++ & JBHI    & \href{https://github.com/DebeshJha/ResUNetPlusPlus-with-CRF-and-TTA}{PyTorch} &  IS, VS     & PraNet and AM & ResUNet++, CRF and TTA\\
   31 &NanoNet & CBMS   & \href{https://github.com/DebeshJha/NanoNet}{PyTorch} &  IS, VS     & ED, KS, and KCS & Real-Time polyp segmentation\\
   32 &ColonSegNet  & Access  & \href{https://github.com/DebeshJha/ColonSegNet}{PyTorch} & IS   & KS   &  Residual block and SENet\\
    33&Segtran & IJCAI  & \href{https://github.com/askerlee/segtran}{PyTorch}&   IS    & C6 and KS& Transformer\\
    34&DDANet & ICPR   & 
    \href{https://github.com/nikhilroxtomar/DDANet }{PyTorch}&    IS   & KS & Dual decoder attention network\\
    35&UACANet& ACM MM   & \href{https://github.com/plemeri/UACANet}{PyTorch}    &       IS &PraNet  &\makecell[l]{ Uncertainty augmented \\Context attention network} \\ 
    36&DivergentNet & ISBI   & \href{https://github.com/vlbthambawita/divergent-nets}{PyTorch}    &       IS & EndoCV 2021 &Combine multiple models\\
    37&DWHieraSeg & MIA  &  \href{https://github.com/CityU-AIM-Group/DW-HieraSeg}{PyTorch}    &       IS & ES &Dynamic-weighting\\
    38&Transfuse & MICCAI   & \href{https://github.com/Rayicer/TransFuse}{PyTorch}  &IS       & PraNet & Transformer and CNN\\
   39 &SANet & MICCAI & \href{https://github.com/weijun88/SANet}{PyTorch} & IS & PraNet & Shallow attention network \\
   40 &PNS-Net & MICCAI  & \href{https://github.com/GewelsJI/PNS-Net}{PyTorch}&   VS    & C6, KS, ES, and AM &\makecell[l]{Progressively normalized \\ self-attention network}
    \\
    \hline
    \end{tabular}%
  \label{tab: model_survey}%
\end{table*}%

\begin{figure*}[t!]
	\centering
    \small
	\begin{overpic}[width=.95\linewidth]{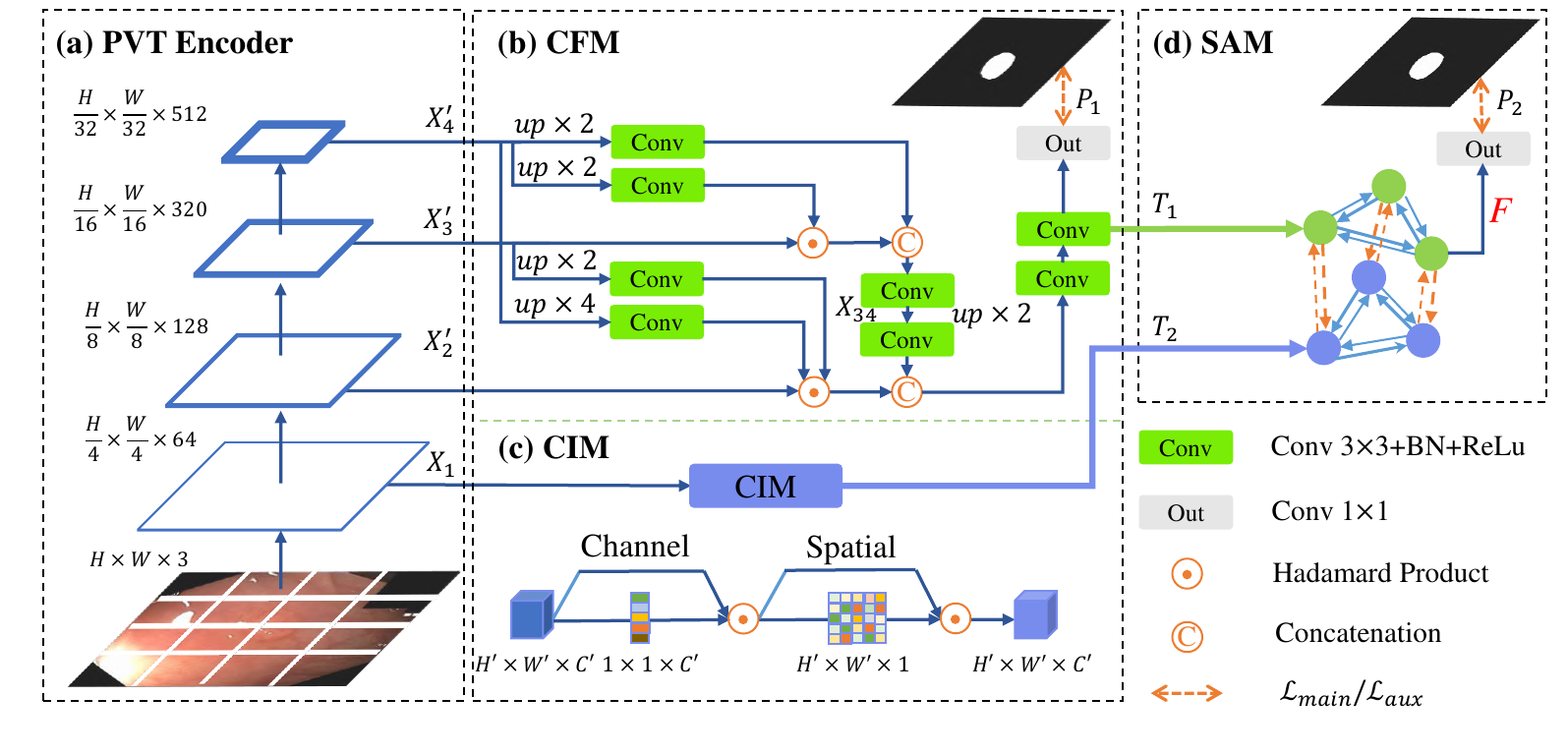}
    \end{overpic}
    \vspace{-20pt}
	\caption{Framework of our \ourmodel, which consists of a pyramid vision transformer (PVT) (a) as the encoder network, (b) cascaded fusion module (CFM) for fusing the high-level feature, (c) camouflage identification module (CIM) to filter out the low-level information, and (d) similarity aggregation module (SAM) for integrating the high- and low-level features for the final output.}
    \label{figure:framework}
\end{figure*}

\subsection{Vision Transformer}\label{subsec:Transformer}
Transformers use multi-head self-attention (MHSA) layers to model long-term dependencies. 
Unlike the convolutional layer, the MHSA layer has dynamic weights and a global receptive field, making it more flexible and effective.
The transformer~\cite{vaswani2017attention} was first proposed by Vaswani \etal~for the machine translation task and has since extensively influenced the natural language processing field.
To apply transformers to computer vision tasks, Dosovitskiy \etal~\cite{dosovitskiy2020image} proposed a vision transformer (ViT), which was the first pure transformer for image classification.
ViT divides an image into multiple patches, which are sequentially sent to a transformer encoder after being encoded, and then an MLP is used to perform image classification. 
HVT~\cite{pan2021scalable} is based on a hierarchical progressive pooling method to compress the sequence length of a token and reduce the redundancy and number of calculations in ViT. 
The pooling-based vision transformer~\cite{heo2021rethinking} draws on the principle of CNNs whereby, as the depth increases, the number of feature map channels increases, and the spatial dimension decreases.
Yuan~\etal~\cite{yuan2021tokens} pointed out that the simple token structure in ViT cannot capture important local features, such as edges and lines, which reduces the training efficiency and leads to redundant attention mechanisms.
T2T ViT was thus proposed to use layer-by-layer tokens-to-token transformation to gradually merge neighboring tokens and model local features while reducing the token's length. 
TNT~\cite{han2021transformer} employs a transformer suitable for fine-grained image tasks, which divides the original image patch and conducts self-attention mechanism calculations in smaller units. 
Meanwhile, external and internal transformers are used to extract global and local features. 

To adapt to dense prediction tasks such as semantic segmentation, 
several methods~\cite{wang2021pyramid,wang2021pvtv2,liu2021swin,cvt,coat,twins,levit} have also introduced the pyramid structure of CNNs to the design of transformer backbones.
For instance, PVT-based models~\cite{wang2021pyramid,wang2021pvtv2} use a hierarchical transformer with four stages, showing that a pure transformer backbone can be as versatile as its CNN counterparts, and performs better in detection and segmentation tasks.
In this work, {we design a new transformer-based polyp segmentation framework}, which can accurately locate the boundaries of polyps even in extreme scenarios.

\section{Proposed Polyp-PVT}\label{sec:Method}
\subsection{Overall Architecture}
As shown in \figref{figure:framework}, our
\ourmodel~consists of 4 key modules: namely, a pyramid vision transformer (PVT) encoder, cascaded fusion module (CFM), camouflage identification module (CIM), and similarity aggregation module (SAM).
Specifically, the PVT extracts multi-scale long-range dependencies features from the input image.
The CFM is employed to collect semantic cues and locate polyps by aggregating high-level features progressively.
The CIM is designed to remove noise and enhance low-level representation information of polyps, including texture, color, and edges.
The SAM is adopted to fuse the low- and high-level features provided by the CIM and CFM, effectively transmitting the information from the pixel-level polyp to the entire polyp.

Given an input image $I \in \mathbb{R}^{H \times W \times 3}$, we use the transformer-based backbone~\cite{wang2021pyramid} to extract four pyramid features $X_{i} \in \mathbb{R}^{\frac{H}{2^{i+1}} \times \frac{W}{2^{i+1}} \times C_i}$, where $C_i \in \{64,128,320,512\}$ and $i\in\{1,2,3,4\}$.
Then, we adjust the channel of three high-level features $X_2$, $X_3$ and $X_4$ to 32 through three convolutional units and feed them (\emph{i.e.}, $X_2^{'}$, $X_3^{'}$, and $X_4^{'}$) to CFM to fuse, leading a feature map $T_1 \in \mathbb{R}^{\frac{H}{8} \times \frac{W}{8} \times 32}$.
Meanwhile, low-level features $X_{1}$ are converted to $T_2 \in \mathbb{R}^{\frac{H}{4} \times \frac{W}{4} \times 64}$ by the CIM.
After that, the $T_1$ and $T_2$ are 
aligned and fused
by SAM, yielding the final feature map $F \in \mathbb{R}^{\frac{H}{8} \times \frac{W}{8} \times 32}$.
Finally, $F$ is fed into a $1\times1$ convolutional layer to predict the polyp segmentation result $P_2$. We use the sum of $P_1$ and $P_2$ as the final prediction.
During training, we optimize the model with a main loss $\mathcal{L}_{\rm main}$ and an auxiliary loss $\mathcal{L}_{\rm aux}$. The main loss is calculated between the final segmentation result $P_2$ and the ground truth (GT), which is used to optimize the final polyp segmentation result.
Similarly, the auxiliary loss is used to supervise the intermediate result $P_1$ generated by the CFM.

\subsection{Transformer Encoder}\label{subsec:Transformer Encoder}
Due to uncontrolled factors in their acquisition, polyp images tend to contain significant noise, such as \textit{motion blur}, \textit{rotation}, and \textit{reflection}.
Some recent works~\cite{bhojanapalli2021understanding,xie2021segformer} have found that the vision transformer~\cite{dosovitskiy2020image,wang2021pyramid,wang2021pvtv2} demonstrates stronger performance and better robustness to input disturbances than CNNs~\cite{simonyan2015very,he2016deep}.
Inspired by this, we use a vision transformer as our backbone network to extract more robust and powerful features for polyp segmentation. 
Different from~\cite{liu2021swin,dosovitskiy2020image} that uses a fixed ``columnar'' structure or shifted windowing manner, the PVT~\cite{wang2021pyramid} is a pyramid architecture whose representation is calculated with spatial-reduction attention operations; thus it enables to reduce the resource consumption.
Note that the proposed model is backbone-independent; other famous transformer backbones are feasible in our framework. Specifically, we adopt the PVTv2~\cite{wang2021pvtv2}, which is the improved version of PVT with a more powerful feature extraction ability.
To adapt PVTv2 to the polyp segmentation task, we remove the last classification layer and design a polyp segmentation head on top of four multi-scale feature maps (\emph{i.e.}, $X_1$, $X_2$, $X_3$, and $X_4$) generated by different stages. 
Among these feature maps, $X_1$ gives detailed appearance information of polyps, and $X_2$, $X_3$, and $X_4$ provide high-level features.

\subsection{Cascaded Fusion Module}\label{subsec:cascaded fusion}
To balance the accuracy and computational resources, we follow recent popular practices~\cite{Wu_2019_CVPR,fan2020pranet} to implement the cascaded fuse module (CFM). 
Specifically, we define $\mathcal{F}(\cdot)$ as a convolutional unit composed of a $3\times 3$ convolutional layer with padding set to 1, batch normalization~\cite{ioffe2015batch} and ReLU~\cite{glorot2011deep}.
As shown in Fig. \ref{figure:framework} (b), the CFM mainly consists of two cascaded parts, as follows:

(1) In part one,
we up-sample the highest-level feature map $X_{4}^{'}$ to the same size as $X_{3}^{'}$ and then pass the result through two convolutional units {$\mathcal{F}_1(\cdot)$ and $\mathcal{F}_2(\cdot)$}, yieldings: $X_{4}^1$ and $X_{4}^2$.
Then, we multiply $X_{4}^1$ and $X_{3}^{'}$ and concatenate the result with $X_{4}^2$.
Finally, we use a convolution unit $\mathcal{F}_3(\cdot)$ to smooth the concatenated feature, yielding fused feature map $X_{34} \in \mathbb{R}^{\frac{H}{16} \times \frac{W}{16} \times 32}$.
The process can be summarized as Eqn.~\ref{eqn:part_1}.
\begin{equation}
X_{34} = \mathcal{F}_{3}(\text{Concat}(\mathcal{F}_1({{{X}}_4^{'}}) \odot {X_{3}^{'}}, \mathcal{F}_2({{{X}}_4^{'}}))),
\label{eqn:part_1}
\end{equation}
where ``$\odot$'' denotes the Hadamard product, and $\text{Concat}(\cdot)$ is the concatenation operation along the channel dimension.

(2) As shown Eqn.~\ref{eqn:part_2}, the second part follows a similar process to part one. 
Firstly, we up-sample $X_{4}^{'}$, $X_{3}^{'}$, $X_{34}$
to the same size as $X_{2}^{'}$, and smooth them using convolutional units $\mathcal{F}_{4}(\cdot)$, $\mathcal{F}_{5}(\cdot)$, and $\mathcal{F}_{6}(\cdot)$, respectively.
Then, we multiply the smoothed $X_{4}^{'}$ and $X_{3}^{'}$ with $X_2^{'}$, and concatenate the resulting map with 
up-sampled and smoothed $X_{34}$.
Finally, we feed the concatenated feature map into two convolutional units (\emph{i.e.}, $\mathcal{F}_{7}(\cdot)$ and $\mathcal{F}_{8}(\cdot)$) to reduce the dimension, and obtain $T_{1} \in \mathbb{R}^{\frac{H}{8} \times \frac{W}{8} \times 32}$, which is also the output of the CFM.
\begin{equation}
T_{1} = \mathcal{F}_{8}(\mathcal{F}_{7}(\text{Concat}(\mathcal{F}_{4}(X_4^{'}) \odot \mathcal{F}_{5}(X_3^{'}) \odot {X_2^{'}}, \mathcal{F}_{6}(X_{34})))),
\label{eqn:part_2}
\end{equation}

\subsection{Camouflage Identification Module}\label{subsec:Camouflage Identification}
Low-level features often contain rich detail 
information, such as \textit{texture}, \textit{color}, and \textit{edges}.
However, polyps tend to be very similar in appearance to the background.
Therefore, we need a powerful extractor to identify the polyp details.

As shown in Fig.~\ref{figure:framework} (c), we introduce a camouflage identification module (CIM) to capture the details of polyps from different dimensions of the low-level feature map $X_1$.
Specifically, the CIM consists of a channel attention operation~\cite{woo2018cbam} $\text{Att}_{c}(\cdot)$ and a spatial attention operation~\cite{hu2018squeeze} $\text{Att}_{s}(\cdot)$, which can be formulated as:
\begin{equation}
T_{2} = \text{Att}_{s}\left(\text{Att}_{c}\left(X_1\right)\right),
\label{equ:att}
\end{equation}
The channel attention operation $\text{Att}_{c}(\cdot)$ can be written as follow:
\begin{equation}\label{equ:loss function}
\text{Att}_{c}(x) = \sigma\left(\mathcal{H}_1\left({P_{\rm max}}\left(x\right)\right) + \mathcal{H}_2\left({P_{\rm avg}}\left(x\right)\right)\right) \odot x,
\end{equation}
where $x$ is the input tensor and $\sigma(\cdot)$ is the Softmax function. $P_{\rm max}(\cdot)$ and $P_{\rm avg}(\cdot)$ denote adaptive maximum pooling and adaptive average pooling functions, respectively.
${\mathcal{H}_{i}(\cdot)}, i \in \{1,2\}$ shares parameters and consists of a convolutional layer with $1 \times 1$ kernel size to reduce the channel dimension 16 times, followed by a ReLU layer
and another $1 \times 1$ convolutional layer to recover the original channel dimension.
The spatial attention operation $\text{Att}_{s}(\cdot)$ can be formulated as:
\begin{equation}
\text{Att}_{s}(x) = \sigma(\mathcal{G}(\text{Concat} ({R_{\rm max}}(x), {R_{\rm avg}}(x)))) \odot x,
\label{equ:att_s}
\end{equation}
where $R_{\rm max}(\cdot)$ and $R_{\rm avg}(\cdot)$ represent the maximum and average values obtained along the channel dimension, respectively.
$\mathcal{G}(\cdot)$ is a $7 \times 7$ convolutional layer with padding set to {3}.

\begin{figure}[t!]
	\centering
    \small
    \vspace{5pt}
	\begin{overpic}[width=.98\linewidth]{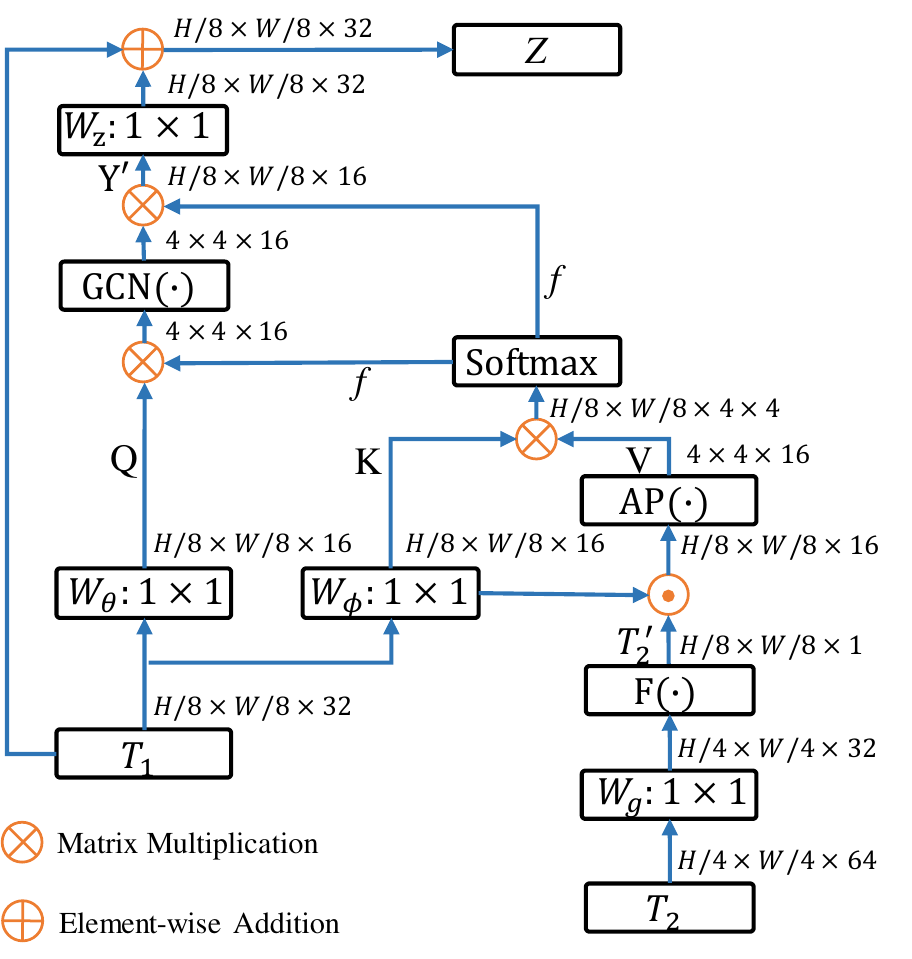}
    \end{overpic}
    \vspace{-10pt}
	\caption{Details of the introduced SAM. It is composed of GCN and non-local, which extend the pixel features of polyp regions with high-level semantic location cues to the entire region.}
    \label{figure:module ra}
    \vspace{8pt}
\end{figure}

\subsection{Similarity Aggregation Module}\label{subsec:GCN}
To explore high-order relations between the lower-level local features from CIM and higher-level cues from CFM. We introduce the non-local~\cite{wang2018non,te2020edge} operation under graph convolution domain~\cite{lu2019graph} to implement our similarity aggregation module (SAM). As a result, SAM can inject detailed appearance features into high-level semantic features using global attention.

Given the feature map ${T_1}$, which contains high-level semantic information, and ${T_2}$ with rich appearance details, we fuse them through self-attention.
First, two linear mapping functions $W_\theta(\cdot)$ and $W_\phi(\cdot)$ are applied on ${T_1}$ to reduce the dimension and obtain feature maps $Q \in \mathbb{R}^{\frac{H}{8} \times \frac{W}{8} \times 16}$ and $K \in \mathbb{R}^{\frac{H}{8} \times \frac{W}{8} \times 16}$. 
Here, we take a convolution operation with a kernel size of $1\times1$ as the linear mapping process. 
This process can be expressed as follows:
\begin{equation}\label{equ:q}
Q = W_\theta (T_1),K =  W_\phi (T_1).
\end{equation}

For ${T_2}$, we use a convolutional unit $W_{g}(\cdot)$ to reduce the channel dimension to 32 and interpolate it to the same size as ${T_1}$. 
Then, we apply a Softmax function on the channel dimension and choose the second channel
as the attention map, leading to ${T^{'}_2} \in \mathbb{R}^{\frac{H}{8} \times \frac{W}{8} \times 1}$.
These operations are represented as $\text{F}(\cdot)$ in Fig.~\ref{figure:module ra}. 
Next, we calculate the Hadamard product between $K$ and ${T^{'}_2}$.
This operation assigns different weights to different pixels, increasing the weight of edge pixels.
After that, we use an adaptive pooling operation to reduce the displacement of features and apply a center crop on it to obtain the feature map
$V \in \mathbb{R}^{4 \times 4 \times 16}$.
In summary, the process can be formulated as follows:
\begin{equation}\label{equ:v}
V = \text{AP}(K \odot \text{F}(W_{g} ({T_2}))),
\end{equation}
where $\text{AP}(\cdot)$ denotes the pooling and crop operations.

Then, we establish the correlation between each pixel in $V$ and $K$ through an inner product, which is written as follows:
\begin{equation}\label{equ:f}
f= \sigma (V^\mathsf{T}  \otimes K),
\end{equation}
where "$\otimes$" denotes the inner product operation. $ V^\mathsf{T}$ is the transpose of $V$ and $f$ is the correlation attention map.

After obtaining the correlation attention map $f$, we multiply it with the feature map $Q$, and the result features are fed to the graph convolutional layer~\cite{te2020edge} $\text{GCN}(\cdot)$, leading to $G \in \mathbb{R}^{4 \times4 \times 16}$. 
Same to~\cite{te2020edge}, we calculate the inner product between $f$ and $G$ as Eqn.~\ref{equ:y}, reconstructing the graph domain features into the original structural features:
\begin{equation}\label{equ:y}
Y'= f^\mathsf{T} \otimes \text{GCN}(f^\mathsf{T}  \otimes Q).
\end{equation}

The reconstructed feature map $Y'$ is adjusted to the same channel sizes with $Y$ by a convolutional layer $W_z(\cdot)$ with $1 \times 1$ kernel size, and then combined with the feature ${T_1}$ to obtain the final output $Z \in \mathbb{R}^{\frac{H}{8} \times \frac{W}{8} \times 32}$ of the SAM. 
Eqn.~\ref{equ:z} summarizes the details of this process:
\begin{equation}\label{equ:z}
Z = T_1 + W_z(Y').
\end{equation}

\begin{figure}[t!]
 	\centering
 	\begin{overpic}[width=\linewidth]{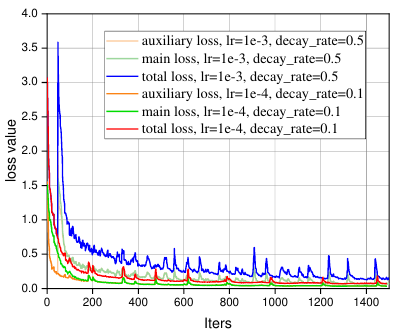}
    \end{overpic}
    \vspace{-10pt}
 	\caption{Loss curves under different training parameter settings.}
     \label{figure:loss}
\end{figure}
\subsection{Loss Function}\label{subsec:Supervision Strategy}

Our loss function can be formulated as Eqn.~\ref{equ:loss}:
\begin{equation}
\mathcal{L} = \mathcal{L}_{\rm main} + \mathcal{L}_{\rm aux},
\label{equ:loss}
\end{equation}
where $\mathcal{L}_{\rm main}$ and $\mathcal{L}_{\rm aux}$ are the main loss and auxiliary loss, respectively. 
The main loss $\mathcal{L}_{\rm main}$ is calculated between the final segmentation result $P_2$ and ground truth $G$, which can be written as:
\begin{equation}
\mathcal{L}_{\rm main} = \mathcal{L}^{w}_{\text{IoU}}(P_2, G) + \mathcal{L}^{w}_{\text{BCE}}(P_2, G).
\label{equ:loss_main}
\end{equation}

The auxiliary loss $\mathcal{L}_{\rm aux}$ is calculated between the intermediate result $P_1$ from the CFM and ground truth $G$, which can be formulated as:
\begin{equation}
\mathcal{L}_{\rm aux} = \mathcal{L}^{w}_{\text{IoU}}(P_1, G) + \mathcal{L}^{w}_{\text{BCE}}(P_1, G).
\label{equ:loss_aux}
\end{equation}

$\mathcal{L}^{w}_{\text{IoU}}(\cdot)$ and $\mathcal{L}^{w}_{\text{BCE}}(\cdot)$ are the weighted intersection over union (IoU) loss \cite{wei2020f3net} and weighted binary cross entropy (BCE) loss \cite{wei2020f3net}, which restrict the prediction map in terms of the global structure (object-level) and local details (pixel-level) perspectives.
Unlike the standard BCE loss function, which treats all pixels equally, $\mathcal{L}^{w}_{\text{BCE}}(\cdot)$ considers the importance of each pixel and assigns higher weights to hard pixels.
Furthermore, compared to the standard IoU loss, $\mathcal{L}^{w}_{\text{IoU}}(\cdot)$ pays more attention to the hard pixels.

\subsection{Implementation Details}\label{subsection: Implementation Details}
We implement our \ourmodel~with the PyTorch framework and use a Tesla P100 to accelerate the calculations.
Considering the differences in the sizes of each polyp image, 
we adopt a multi-scale strategy~\cite{fan2020pranet,huang2021hardnet} in the training stage. 
The hyperparameter details are as follows.
To update the network parameters, we use the AdamW~\cite{loshchilov2019adamw} optimizer, which is widely used in transformer networks~\cite{wang2021pyramid,liu2021swin,wang2021pvtv2}. 
The learning rate is set to 1e-4 and the weight decay is adjusted to 1e-4 too. 
Further, we resize the input images to $352\times352$ with a mini-batch size of 16 for 100 epochs. More details about the training loss cures, parameter setting, and network parameters are shown in \figref{figure:loss}, \tabref{tab:train parameters}, and \tabref{tab:net_paras}, respectively.
The total training time is nearly 3 hours to achieve the best (\emph{e.g.}, 30 epochs) performance.
For testing, we only resize the images to $352\times352$ without any post-processing optimization strategies.

\begin{table}

\small
  \centering
  \caption{Parameter setting during the training stage.}
   \renewcommand{\arraystretch}{1.0}
   {
	\setlength\tabcolsep{6.0pt}
    \begin{tabular}{cccc}
    \hline
    Optimizer & Learning Rate (lr) & Multi-scale & Clip\\
     AdamW & 1e-4 & [0.75,1,1.25] & 0.5\\
    \hline
     Decay rate & Weight decay & Epochs & Input Size \\
     0.1 & 1e-4 & 100   & $352\times352$ \\
    \hline
    \end{tabular}%
  \label{tab:train parameters}}%
  \vspace{8pt}
\end{table}%

\section{Experiments}\label{sec:Experiments}
\subsection{Evaluation Metrics} \label{sec:Metrics}
We employ six widely-used evaluation metrics, including Dice~\cite{milletari2016v}, IoU, mean absolute error (MAE), weighted F-measure (${ F}_{\beta }^{w}$){~\cite{margolin2014evaluate}}, S-measure (${S}_{\alpha}$)~\cite{Fan2021S-measure}, and E-measure (${E}_{\xi}$)~\cite{21Fan_HybridLoss,Fan2018Emeasure}  to evaluate the model performances.
Among these metrics, Dice and IoU are similarity measures at the regional level, which mainly focus on the internal consistency of segmented objects. 
Here, we report the mean value of Dice and IoU, denoted as mDic and mIoU, respectively. 
MAE is a pixel-by-pixel comparison indicator that represents the average value of the absolute error between the predicted value and the true value. 
Weighted F-measure (${ F}_{\beta }^{w}$) comprehensively considers the recall and precision and eliminates the effect of considering each pixel equally in conventional indicators.
S-measure (${S}_{\alpha}$) focuses on the structural similarity of target prospects at the region and object level.
E-measure (${E}_{\xi}$) is used to evaluate the segmentation results at the pixel and image level.
We report the mean and max value of E-measure, denoted as $m{E}_{\xi}$ and $max{E}_{\xi}$, respectively.
The evaluation toolbox is derived from 
\url{https://github.com/DengPingFan/PraNet}.

\begin{table}
\small
  \centering
  \caption{Network parameters of each module. Note that the encoder parameters are the same as PVT without any changes. BasicConv2d and Conv2d with the parameters [in\_channel, out\_channel, kernel\_size, padding] and GCN [num\_state, num\_node].}
  \setlength\tabcolsep{2.5pt}
    \begin{tabular}{lc|lc}
    \hline
    \multicolumn{2}{c|}{\textbf{Encoder}} & \multicolumn{2}{c}{\textbf{SAM}} \\
    \hline
    patch\_size & [4]     & AvgPool2d & [6] \\
    embed\_dims & [64, 128, 320, 512] & Conv2d & [32,16,1,1] \\
    num\_heads & [1, 2, 5, 8] & Conv2d & [32,16,1,1] \\
    mlp\_ratios & [8, 8, 4, 4] & Conv2d & [16,32,1,1] \\
    depths & [3, 4, 18, 3] & GCN   & [16,16] \\
    sr\_ratios & [8, 4, 2, 1] & BasicConv2d & [64,32,1,0] \\
    drop\_rate & [0]     &       &  \\
    drop\_path\_rate & [0.1]   &       &  \\
    \hline
    \multicolumn{2}{c|}{\textbf{CFM}} & \multicolumn{2}{c}{\textbf{CIM}} \\
    \hline
    BasicConv2d & [32,32,3,1] & AvgPool2d & [1] \\
    BasicConv2d & [32,32,3,1] & AvgPool2d & [1] \\
    BasicConv2d & [32,32,3,1] & Conv2d & [64,4,1,0] \\
    BasicConv2d & [32,32,3,1] & ReLU  &  \\
    BasicConv2d & [64,64,3,1] & Conv2d & [4,64,1,0] \\
    BasicConv2d & [64,64,3,1] & Sigmoid &  \\
    BasicConv2d & [96,96,3,1] & Conv2d & [2,1,7,3] \\
    BasicConv2d & [96,32,3,1] & Sigmoid &  \\
    \hline
    \end{tabular}%
  \label{tab:net_paras}%
\end{table}%

\subsection{Datasets and Compared Models}
\textcolor[RGB]{31,100,212}{Datasets.} Following the experimental setups in PraNet~\cite{fan2020pranet}, we adopt five challenging public datasets, including Kvasir-SEG~\cite{jha2020kvasir}, ClinicDB~\cite{bernal2015wm}, ColonDB~\cite{tajbakhsh2015automated}, Endoscene~\cite{vazquez2017benchmark} and ETIS~\cite{silva2014toward} to verify the effectiveness of our framework.

\textcolor[RGB]{31,100,212}{Models.} We collect several open source models from the field of polyp segmentation, for a total of nine comparative models, including U-Net~\cite{ronneberger2015unet}, UNet++~\cite{zhou2018unet++}, PraNet~\cite{fan2020pranet}, SFA~\cite{fang2019sfa}, MSEG~\cite{huang2021hardnet}, ACSNet~\cite{zhang2020ACSN}, DCRNet~\cite{yin2021duplex}, EU-Net~\cite{PatelBW21} and SANet~\cite{wei2021shallow}. 
For a fair comparison, we use their open-source codes to evaluate the same training and testing sets. Note that the SFA results are generated using the released test model. 

\subsection{Analysis of Learning Ability}
\textcolor[RGB]{31,100,212}{Settings.} We use the ClinicDB and Kvasir-SEG datasets to evaluate the learning ability of the proposed model. ClinicDB contains 612 images, which are extracted from 31 colonoscopy videos. Kvasir-SEG is collected from the polyp class in the Kvasir dataset and includes 1,000 polyp images.
Following PraNet, we adopt the same 900 and 548 images from ClinicDB and Kvasir-SEG datasets as the training set, and the remaining 64 and 100 images are employed as the respective test sets.

\begin{table*}[t!]
\small
\centering
	\caption{Quantitative results of the test datasets, \emph{i.e.}, Kvasir-SEG and ClinicDB.
	}
	\vspace{-5pt}
	\renewcommand{\arraystretch}{1.0}
	\setlength\tabcolsep{4.0pt}
    \begin{tabular}{l|ccccccc|ccccccc}
    \hline  
          & \multicolumn{7}{c|}{Kvasir-SEG~\cite{jha2020kvasir}}                                  & \multicolumn{7}{c}{ClinicDB~\cite{bernal2015wm}} \\
\cline{2-15} 
Model & mDic& mIoU & ${ F}_{\beta }^{w}$   & ${S}_{\alpha}$    & $m{E}_{\xi}$   & $max{E}_{\xi}$  & MAE & mDic & mIoU & ${ F}_{\beta }^{w}$   & ${S}_{\alpha}$    & $m{E}_{\xi}$ & $max{E}_{\xi}$  & MAE   \\
    \hline
    MICCAI'15 U-Net  & 0.818  & 0.746  & 0.794  & 0.858  & 0.881    & 0.893 & 0.055  & 0.823  & 0.755  & 0.811  & 0.889  & 0.913    & 0.954 & 0.019    \\
    DLMIA'18 UNet++  & 0.821  & 0.743  & 0.808  & 0.862  & 0.886    & 0.909  & 0.048  & 0.794  & 0.729  & 0.785  & 0.873  & 0.891   & 0.931 & 0.022    \\
    MICCAI'19 SFA     & 0.723  & 0.611  & 0.670  & 0.782  & 0.834    & 0.849& 0.075  & 0.700  & 0.607  & 0.647  & 0.793  & 0.840   & 0.885  & 0.042  \\
    arXiv'21 MSEG & 0.897  & 0.839  & 0.885  & 0.912  & 0.942   & 0.948  & 0.028   & 0.909  & 0.864  & 0.907  & 0.938  & 0.961    & 0.969 & 0.007    \\
    arXiv'21 DCRNet  & 0.886  & 0.825  & 0.868  & 0.911  & 0.933   & 0.941 & 0.035  & 0.896  & 0.844  & 0.890  & 0.933  & 0.964   & 0.978 & 0.010   \\
    MICCAI'20 ACSNet  & 0.898  & 0.838  & 0.882  & 0.920  & 0.941   & 0.952   & 0.032   & 0.882  & 0.826  & 0.873  & 0.927  & 0.947   & 0.959 & 0.011   \\
    MICCAI'20 PraNet  & 0.898  & 0.840  & 0.885  & 0.915  & 0.944    & 0.948 & 0.030   & 0.899  & 0.849  & 0.896  & 0.936  & 0.963   & 0.979 & 0.009    \\
    CRV'21 EU-Net & 0.908  & 0.854  & 0.893  & 0.917  & 0.951  & 0.954  & 0.028    & 0.902  & 0.846  & 0.891  & 0.936  & 0.959   & 0.965 & 0.011    \\
    MICCAI'21 SANet & 0.904  & 0.847  & 0.892  & 0.915  & 0.949    & 0.953  & 0.028 & 0.916  & 0.859  & 0.909  & 0.939  & 0.971    & 0.976 & 0.012   \\
    \hline
    \rowcolor{gray!30}
    \textbf{\ourmodel~(Ours)} & \textbf{0.917} & \textbf{0.864} & \textbf{0.911} & \textbf{0.925} & \textbf{0.956}  & \textbf{0.962} & \textbf{0.023}  & \textbf{0.937} & \textbf{0.889} & \textbf{0.936} & \textbf{0.949} & \textbf{0.985}  & \textbf{0.989} & \textbf{0.006}\\
    \hline
    \end{tabular}
  \label{tab:test1_1}
\end{table*}%

\begin{table*}[t!]
\small
    \centering
    \renewcommand{\arraystretch}{1.0}
	\caption{Quantitative results of the test datasets ColonDB and ETIS. The SFA result is generated using the published code.}
	\vspace{-5pt}
    \setlength\tabcolsep{4.0pt}
    \begin{tabular}{l|ccccccc|ccccccc}
    \hline
    & \multicolumn{7}{c|}{ColonDB~\cite{tajbakhsh2015automated}}                                  & \multicolumn{7}{c}{ETIS~\cite{silva2014toward}} \\
    \cline{2-15}    
    Model & mDic & mIoU & ${ F}_{\beta }^{w}$   & ${S}_{\alpha}$    & $m{E}_{\xi}$   & $max{E}_{\xi}$ & MAE  & mDic & mIoU & ${ F}_{\beta }^{w}$   & ${S}_{\alpha}$    & $m{E}_{\xi}$  & $max{E}_{\xi}$  & MAE  \\
    \hline
    MICCAI'15 U-Net  & 0.512  & 0.444  & 0.498  & 0.712  & 0.696   & 0.776  & 0.061  & 0.398  & 0.335  & 0.366  & 0.684  & 0.643   & 0.740   & 0.036  \\
    DLMIA'18 UNet++ & 0.483  & 0.410  & 0.467  & 0.691  & 0.680    & 0.760 & 0.064   & 0.401  & 0.344  & 0.390  & 0.683  & 0.629    & 0.776 & 0.035  \\
    MICCAI'19 SFA   & 0.469  & 0.347  & 0.379  & 0.634  & 0.675   & 0.764    & 0.094 & 0.297  & 0.217  & 0.231  & 0.557  & 0.531    & 0.632  & 0.109   \\
    MICCAI'20 ACSNet & 0.716  & 0.649  & 0.697  & 0.829  & 0.839    & 0.851 & 0.039  & 0.578  & 0.509  & 0.530  & 0.754  & 0.737    & 0.764  & 0.059 \\
    arXiv'21 MSEG & 0.735  & 0.666  & 0.724  & 0.834  & 0.859    & 0.875 & 0.038   & 0.700  & 0.630  & 0.671  & 0.828  & 0.854   & 0.890  & 0.015  \\
    arXiv'21 DCRNet & 0.704  & 0.631  & 0.684  & 0.821  & 0.840   & 0.848 & 0.052  & 0.556  & 0.496  & 0.506  & 0.736  & 0.742   & 0.773  & 0.096  \\
    MICCAI'20 PraNet & 0.712  & 0.640  & 0.699  & 0.820  & 0.847   & 0.872  & 0.043   & 0.628  & 0.567  & 0.600  & 0.794  & 0.808   & 0.841 & 0.031    \\
    CRV'21 EU-Net & 0.756  & 0.681  & 0.730  & 0.831  & 0.863    & 0.872 & 0.045  & 0.687  & 0.609  & 0.636  & 0.793  & 0.807    & 0.841 & 0.067 \\
    MICCAI'21 SANet & 0.753  & 0.670  & 0.726  & 0.837  & 0.869   & 0.878  & 0.043  & 0.750  & 0.654  & 0.685  & 0.849  & 0.881   & 0.897   & 0.015 \\
    \hline
    \rowcolor{gray!30}
    \ourmodel~(Ours) & \textbf{0.808} & \textbf{0.727} & \textbf{0.795} & \textbf{0.865} & \textbf{0.913} & \textbf{0.919} & \textbf{0.031}  & \textbf{0.787} & \textbf{0.706} & \textbf{0.750} & \textbf{0.871} & \textbf{0.906}  & \textbf{0.910} & \textbf{0.013} \\
    \hline
    \end{tabular}%
  \label{tab:test1_2}
\end{table*}%

\begin{figure}[t!]
	\centering
	\vspace{5pt}
	\begin{overpic}[width=.8\linewidth]{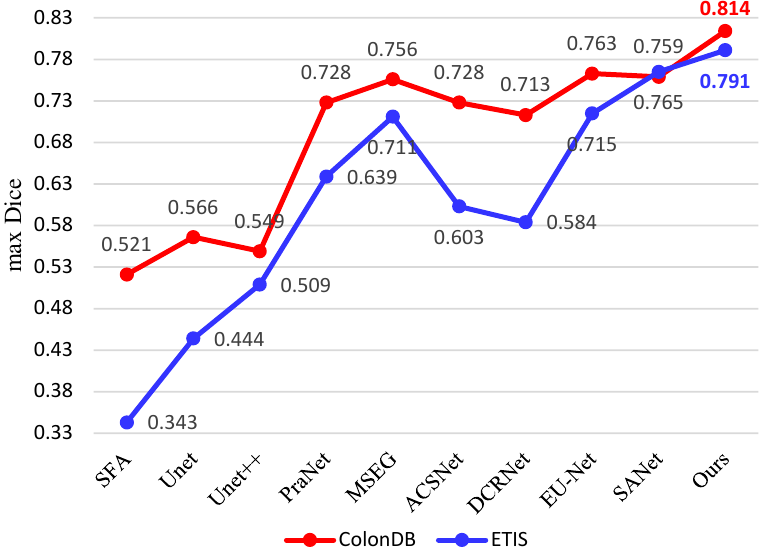}
    \end{overpic}
    \vspace{-5pt}
	\caption{Evaluation of model generalization ability. We provide the max Dice results on ColonDB and ETIS.}
    \label{figure:generalization ability}
\end{figure}

\textcolor[RGB]{31,100,212}{Results.} As can be seen in \tabref{tab:test1_1}, our model is superior to the current methods, demonstrating that it has a better learning ability.
On the Kvasir-SEG dataset, the mDic score of our model is 1.3\% higher than that of the second-best model, SANet, and 1.9\% higher than that of PraNet. 
On the ClinicDB dataset, the mDic score of our model is 2.1\% higher than that of SANet and 3.8\% higher than that of PraNet.

\begin{table}
 \centering
 \small
	\caption{Quantitative results of the test dataset Endoscene. The SFA result is generated using the published code.}\label{tab:test1_3}
	\renewcommand{\arraystretch}{0.8}
	\setlength\tabcolsep{3.0pt}
    \begin{tabular}{l|ccccccc}
    \hline
     & \multicolumn{7}{c}{Endoscene~\cite{vazquez2017benchmark}} \\
\cline{2-8}    Model & mDic & mIoU & ${ F}_{\beta }^{w}$   & ${S}_{\alpha}$    & $m{E}_{\xi}$    & $max{E}_{\xi}$ & MAE \\
    \hline
    U-Net  & 0.710  & 0.627  & 0.684  & 0.843  & 0.847    & 0.875& 0.022    \\
    UNet++ & 0.707  & 0.624  & 0.687  & 0.839  & 0.834  & 0.898  & 0.018   \\
    SFA   & 0.467  & 0.329  & 0.341  & 0.640  & 0.644    & 0.817 & 0.065  \\
    MSEG & 0.874  & 0.804  & 0.852  & 0.924  & 0.948    & 0.957 & 0.009   \\
    ACSNet & 0.863  & 0.787  & 0.825  & 0.923  & 0.939  & 0.968  & 0.013  \\
    DCRNet & 0.856  & 0.788  & 0.830  & 0.921  & 0.943   & 0.960   & 0.010  \\
    PraNet & 0.871  & 0.797  & 0.843  & 0.925  & 0.950   & 0.972 & 0.010    \\
    EU-Net & 0.837  & 0.765  & 0.805  & 0.904  & 0.919    & 0.933  & 0.015  \\
    SANet & 0.888  & 0.815  & 0.859  & 0.928  & 0.962   & 0.972  & 0.008  \\
    \hline
    \rowcolor{gray!30}
    \ourmodel & \textbf{0.900} & \textbf{0.833} & \textbf{0.884} & \textbf{0.935} & \textbf{0.973}& \textbf{0.981}   & \textbf{0.007}  \\
    \hline
    \end{tabular}
\end{table}

\begin{table*}[t!]
  \centering
  \small
  \caption{The standard deviation (SD) of the mean dice (mDic) of our model and the comparison models.}
  \vspace{-5pt}
  \setlength\tabcolsep{16pt}
	\renewcommand{\arraystretch}{1}
    \begin{tabular}{l|c|c|c|c|c}
    \hline 
    Datasets &Kvasir-SEG & ClinicDB &ColonDB & ETIS & Endoscene \\
    \hline 
    Metrics& mDic $ \pm$ SD    & mDic $ \pm$ SD    & mDic $ \pm$  SD    & mDic $ \pm$  SD    & mDic $ \pm$  SD \\
    \hline 
    MICCAI'15 U-Net  & .818 $ \pm$ .039  & .823  $ \pm$ .047  & .483 $ \pm$  .034  & .398  $ \pm$  .033  & .710  $ \pm$  .049  \\
    DLMIA'18 UNet++ & .821  $ \pm$ .040  & .794 $ \pm$ .044  & .456  $ \pm$  .037  & .401  $ \pm$  .057  & .707  $ \pm$  .053  \\
    MICCAI'19 SFA    & .723 $ \pm$  .052  & .701 $ \pm$  .054  & .444  $ \pm$  .037  & .297  $ \pm$  .025  & .468  $ \pm$  .050  \\
    arXiv'21 MSEG  & .897  $ \pm$  .041  & .910  $ \pm$  .048  & .735  $ \pm$  .039  & .700  $ \pm$  .039  & .874  $ \pm$  .051  \\
    MICCAI'20 ACSNet & .898  $ \pm$  .045  & .882  $ \pm$  .048  & .716  $ \pm$  .040  & .578  $ \pm$  .035  & .863  $ \pm$ .055  \\
    arXiv'21 DCRNet & .886  $ \pm$  .043  & .896  $ \pm$  .049  & .704  $ \pm$  .039  & .556  $ \pm$  .039  & .857  $ \pm$  .052  \\
    MICCAI'20 PraNet & .898  $ \pm$  .041  & .899  $ \pm$  .048  & .712  $ \pm$  .038  & .628 $ \pm$  .036  & .871  $ \pm$  .051  \\
    CRV'21 EU-Net & .908 $ \pm$  .042  & .902  $ \pm$  .048  & .756 $ \pm$  .040  & .687  $ \pm$ .039  & .837  $ \pm$  .049  \\
    MICCAI'21 SANet & .904  $ \pm$ .042  & .916  $ \pm$  .049  & .752  $ \pm$  .040  & .750  $ \pm$  .047  & .888  $ \pm$  .054  \\
    \hline 
    \rowcolor{gray!30}
    \ourmodel~(Ours) & \textbf{.917}  $ \pm$  .042  & \textbf{.937}  $ \pm$  .050  & \textbf{.808}  $ \pm$  .043  & \textbf{.787}  $ \pm$  .044  & \textbf{.900} $ \pm$  .052  \\
    \hline 
    \end{tabular}%
  \label{tab:sd}%
\end{table*}%

\subsection{Analysis of Generalization Ability}
\textcolor[RGB]{31,100,212}{Settings.} To verify the generalization performance of the model, we test it on three unseen (\emph{i.e.}, Polycentric) datasets, namely ETIS, ColonDB, and EndoScene.
There are 196 images in ETIS, 380 images in ColonDB, and 60 images in EndoScene.  
It is worth noting that the images in these datasets belong to different medical centers. In other words, the model has not seen their training data, which is different from the verification methods of ClinicDB and Kvasir-SEG.

\begin{figure*}[t!]
	\centering
	\begin{overpic}[width=.96\linewidth]{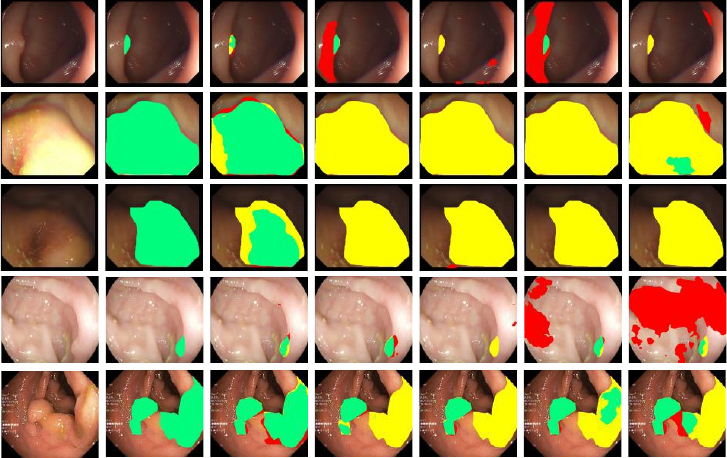}
	\small
	\put(5,-1.5){Image}
	\put(20,-1.5){GT}
	\put(33,-1.5){\color{red}Ours}
	\put(48,-1.5){SANet}
	\put(62,-1.5){PraNet}
	\put(75,-1.5){ACSNet}
	\put(90,-1.5){DCRNet}
    \end{overpic}
	\caption{Visualization results with the current models. Green indicates a correct polyp. Yellow is the missed polyp. Red is the wrong prediction. As we can see, the proposed model can accurately locate and segment polyps, regardless of size.}
    \label{figure:V1}
    \vspace{10pt}
\end{figure*}

\begin{figure*}[t!]
	\centering
	\begin{overpic}[width=.96\linewidth]{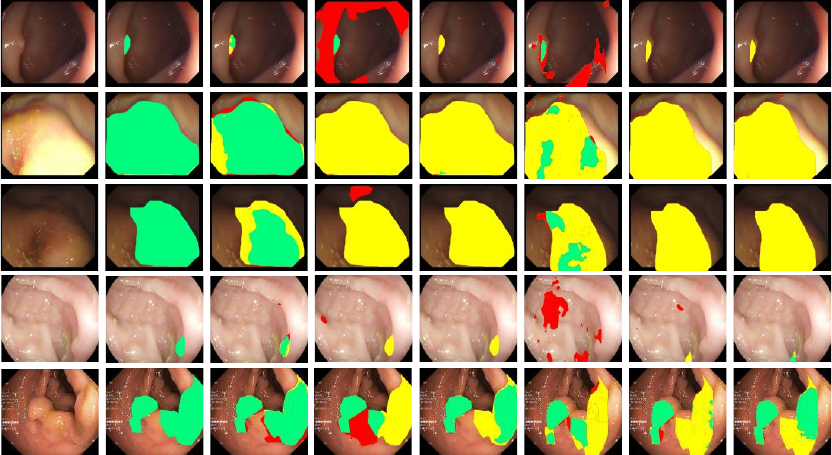}
    \small
	\put(4,-1.5){Image}
	\put(17,-1.5){GT}
	\put(30,-1.5){\color{red}Ours}
	\put(41,-1.5){EU-Net}
	\put(53,-1.5){HarDNet}
	\put(67,-1.5){SFA}
	\put(79,-1.5){U-Net}
	\put(91,-1.5){UNet++}
    \end{overpic}
	\caption{Visualization results with the current models.}
    \label{figure:V2}
\end{figure*}

\textcolor[RGB]{31,100,212}{Results.} The results are shown in Tab.~\ref{tab:test1_3} and Tab.~\ref{tab:test1_2}. As can be seen, our \ourmodel~achieves a good generalization performance compared with the existing models. And our model generalizes easily to multicentric (or unseen) data with different domains/distributions.
On ColonDB, it is ahead of the second-best SANet and classical PraNet by 5.5\% and 9.6\%, respectively. On ETIS, we exceed the SANet and PraNet by 3.7\% and 15.9\%, respectively. In addition, on EndoScene, our model is better than SANet and PraNet by 1.2\% and 2.9\%, respectively. 
Moreover, to prove the generalization ability of \ourmodel, we present the max Dice results in \figref{figure:generalization ability}, where our model shows a steady improvement on both ColonDB and ETIS.
In addition, we show the standard deviation (SD) of the mean dice (mDic) between our model and others in \tabref{tab:sd}. As seen, there is not much difference in SD between our model and the comparison model, and they are both stable and balanced.

\subsection{Qualitative Analysis}

\begin{figure}[t!]
 	\centering
 	\begin{overpic}[width=.99\linewidth]{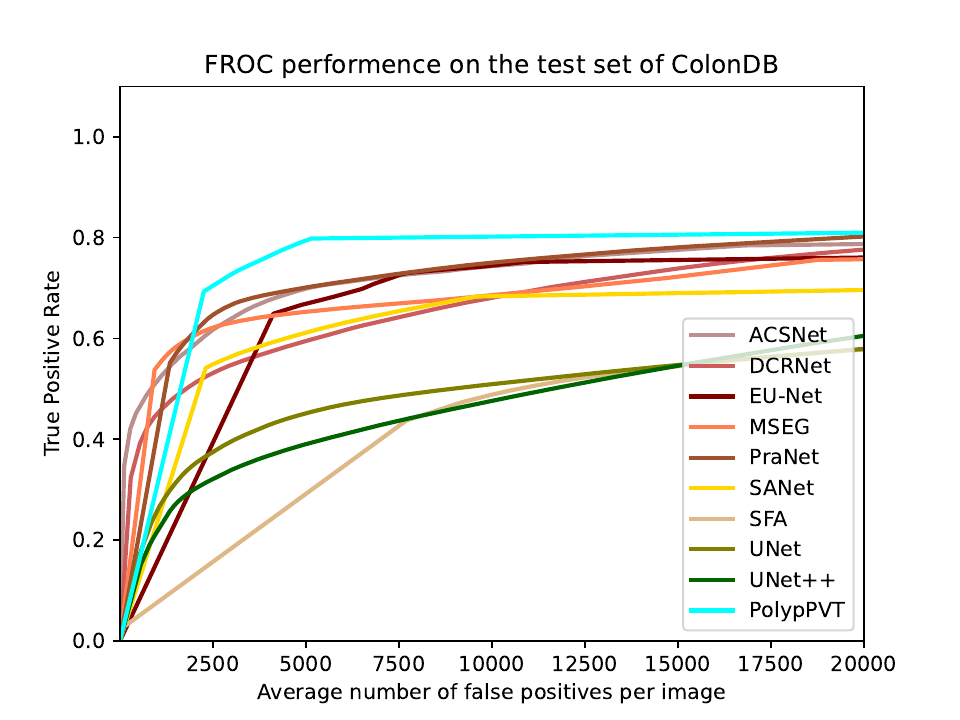}
    \end{overpic}
    \vspace{-10pt}
 	\caption{FROC curves of different methods on ColonDB.}
     \label{figure:froc}
\end{figure}

Fig.~\ref{figure:V1} and Fig.~\ref{figure:V2} show the visualization results of our model and the compared models. We find that our results have two advantages. 
\begin{itemize}
    \vspace{-5pt}
    \item Our model is able to adapt to data under different conditions. That is, it maintains a stable recognition and segmentation ability under different acquisition environments (different lighting, contrast, reflection, motion blur, small objects, and rotation).
    \vspace{-5pt}
    \item The model segmentation results have internal consistency and predicted edges are closer to the ground-truth labels. We also provide FROC curves on ColonDB in Fig.~\ref{figure:froc}, and our result is at the top, indicating that our effect achieves the best. 
    \vspace{-5pt}
\end{itemize}

\subsection{Ablation Study}
We describe in detail the effectiveness of each component on the overall model. 
The training, testing, and hyperparameter settings are the same as mentioned in \secref{subsection: Implementation Details}.
The results are shown in Tab.~\ref{tab:abalation study}.

\begin{figure}[t!]
 	\centering
 	\begin{overpic}[width=\linewidth]{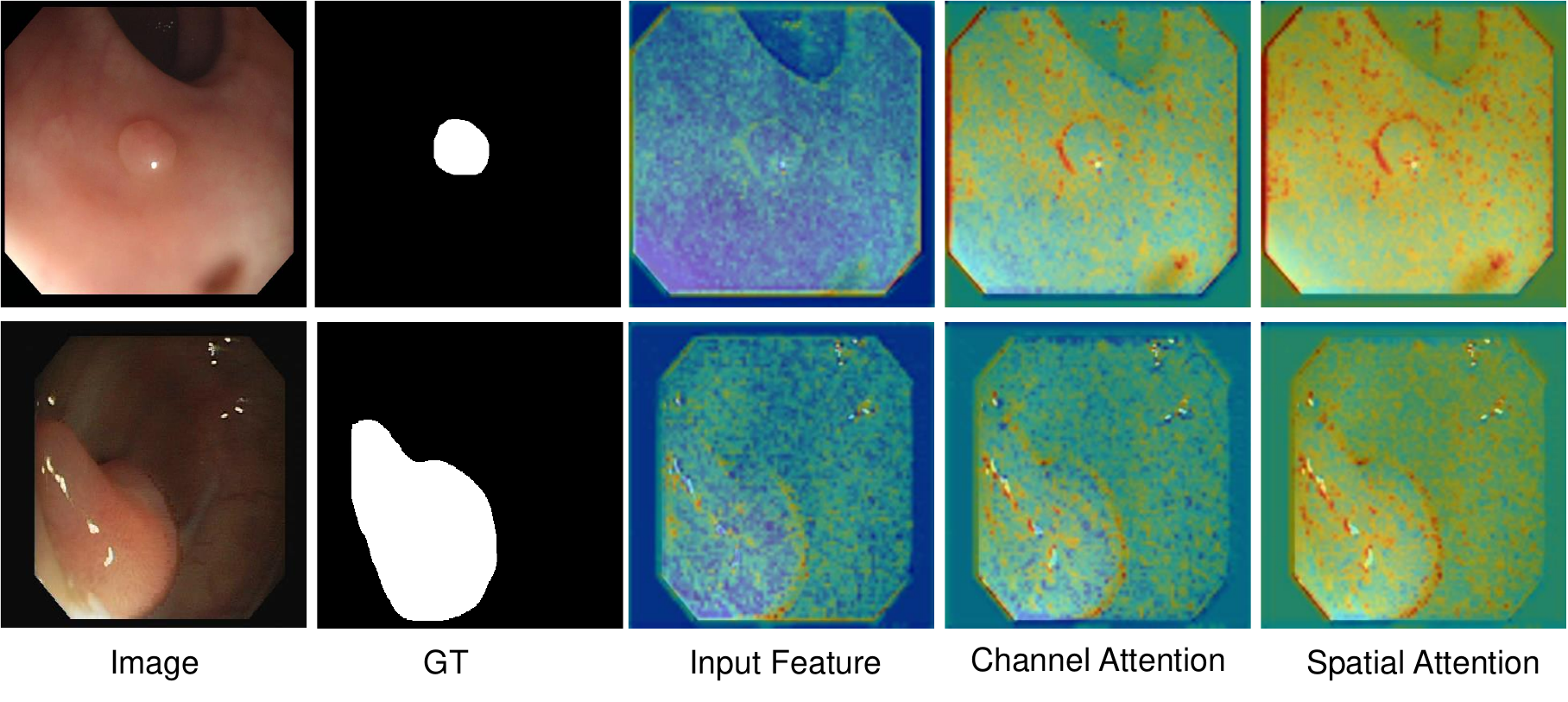}
    \end{overpic}
    \vspace{-10pt}
 	\caption{Visualization of the feature map in the CIM module.}
     \label{figure:cimv}
\end{figure}

\begin{table}
  \centering
  \small
	\caption{Quantitative results for ablation studies.}\label{tab:abalation study}
	\setlength\tabcolsep{1.4pt}
	\renewcommand{\arraystretch}{1.0}
    \begin{tabular}{lc|c|c|c|c|c}
    \hline
    Dataset & Metric & Bas.     & w/o CFM     & w/o CIM      & w/o SAM & Final\\
    \hline
    \multirow{2}[0]{*}{Endoscene} & mDic  & 0.869 & 0.892 & 0.882 & 0.874 & {0.900} \\
          & mIoU  & 0.792 & {0.826} & 0.808 & 0.801 & {0.833} \\
  \hline
    \multirow{2}[0]{*}{ClinicDB} & mDic  & 0.903 & 0.915 & 0.930 & 0.930 & {0.937} \\
          & mIoU  & 0.847 & 0.865 & 0.881& 0.877 & {0.889} \\
\hline
    \multirow{2}[0]{*}{ColonDB} & mDic  & 0.796  & 0.802 & 0.805 & 0.779 & {0.808} \\
          & mIoU  & 0.707   & 0.721 & 0.724 & 0.696 & {0.727} \\
  \hline
    \multirow{2}[0]{*}{ETIS} & mDic  & 0.759 & 0.771 & 0.785 & 0.778 & {0.787} \\
          & mIoU  & 0.668 & 0.690 & {0.711} & 0.693 & {0.706} \\
  \hline
    \multirow{2}[0]{*}{Kvasir-SEG} & mDic  & 0.910 & 0.922 & 0.910 & 0.910 & 0.917 \\
          & mIoU  & 0.856 & 0.872 & 0.858 & 0.853 & {0.864} \\
      \hline
    \end{tabular}
\end{table}%

\textcolor[RGB]{31,100,212}{Components.} 
We use PVTv2~\cite{wang2021pvtv2} as our baseline (Bas.) and evaluate module effectiveness by removing or replacing components from the complete Polyp-PVT and comparing the variants with the standard version. 
The standard version is denoted as ``Polyp-PVT (PVT+CFM+CIM+SAM)'', where ``CFM'', ``CIM'' and ``SAM'' indicate the usage of the CFM, CIM, and SAM, respectively.

\textcolor[RGB]{31,100,212}{Effectiveness of CFM.} To analyze the effectiveness of the CFM, a version of ``Polyp-PVT (w/o CFM)'' is trained. 
\tabref{tab:abalation study} shows that the model without the CFM drops sharply on all five datasets compared to the standard Polyp-PVT.
In particular, the mDic is reduced from 0.937 to 0.915 on ClinicDB.

\begin{figure*}[t!]
	\centering
	\begin{overpic}[width=\linewidth]{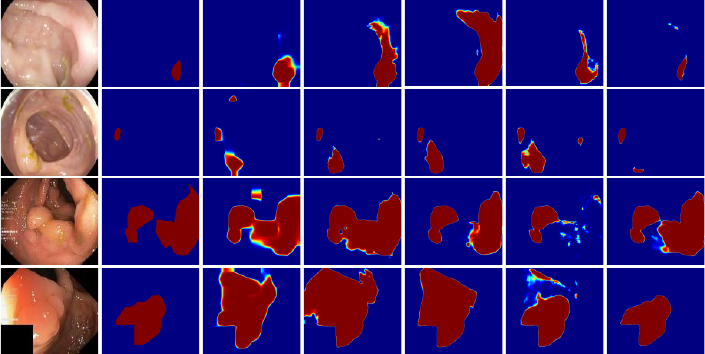}
    \small
	\put(6,-2.5){Image}
	\put(21,-2.5){GT}
	\put(35,-2.5){Bas.}
	\put(46,-2.5){w/o CFM}
	\put(60,-2.5){w/o CIM}
	\put(75,-2.5){w/o SAM}
	\put(91,-2.5){\color{red}Ours}
    \end{overpic}
    \vspace{5pt}
	\caption{Visualization of the ablation study results, which are converted from the output into heat maps. As can be seen, removing any module leads to missed or incorrectly detected results.}
    \label{figure:Visualization_ablation study}
\end{figure*}
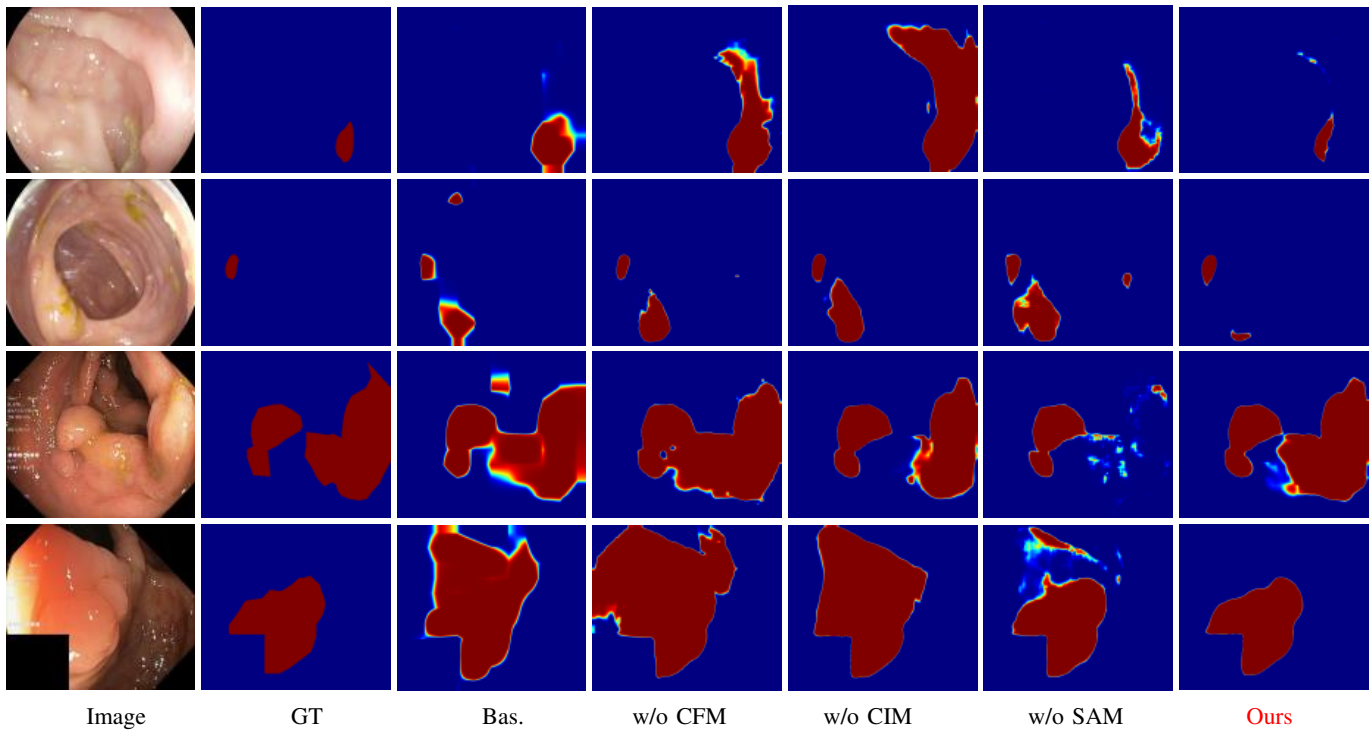

\textcolor[RGB]{31,100,212}{Effectiveness of CIM.} To demonstrate the ability of the CIM, we also remove it from \ourmodel, denoting this as ``Polyp-PVT (w/o CIM)''.
As shown in Tab.~\ref{tab:abalation study}, this variant performs worse than the overall Polyp-PVT. 
Specifically, removing the CIM causes the mDic to decrease by 1.8\% on Endoscene.
Meanwhile, it is obvious that the lack of the CIM introduces significant noise (please refer to Fig.~\ref{figure:Visualization_ablation study}).
In order to further explore the internal of CIM, the feature visualizations of the two main configurations inside the CIM are shown in Fig~\ref{figure:cimv}. It can be seen that the low-level features have a large amount of detailed information. Still, the differences between polyps and other normal tissues cannot be mined directly from this information. Thanks to the channel attention and spatial attention mechanism, information such as details and edges of polyps can be discerned from a large amount of redundant information.

\begin{table}
  \centering
  \caption{Ablation study of GCN in the SAM module.}
  \renewcommand{\arraystretch}{1.1}
  \setlength\tabcolsep{2.1pt}
    \begin{tabular}{l|c|c|c|c|c}
    \hline
     Setting & Endoscene & ClinicDB & ColonDB & ETIS  & Kvasir-SEG \\
    \hline

     w/o GCN & 0.876  & 0.928  & 0.784  & 0.725  & 0.894  \\
           w/ Conv & 0.894  & 0.919  & 0.787  & 0.742  & {0.909}  \\
        w/ GCN & \textbf{0.900} & \textbf{0.937} & \textbf{0.808} & \textbf{0.787} & \textbf{0.917} \\
    \hline
    \end{tabular}\label{tab:gcn}%
\end{table}%

\begin{table}
  \centering
  
  \caption{About the ablation experiments of the powerful rotation adaptability. All experiments are under the condition of large rotation (15 degrees).}
  \renewcommand{\arraystretch}{1.1}
  \setlength\tabcolsep{1.9pt}
    \begin{tabular}{l|c|c|c|c|c}
    \hline
     Setting & Endoscene & ClinicDB & ColonDB & ETIS  & Kvasir-SEG \\
    \hline
               w/o GCN & 0.857  & 0.909  & 0.756  & 0.667  & {0.894}  \\
           w/ Conv & 0.865  & 0.898  & 0.789  & 0.719  & {0.893}  \\
           w/ GCN & \textbf{0.874} & \textbf{0.929} & \textbf{0.806} & \textbf{0.744} & \textbf{0.915} \\
          \hline
    \end{tabular}%
  \label{tab:gcn rotation}%
  \vspace{10pt}
\end{table}%

\begin{figure}[t!]
 	\centering
 	\begin{overpic}[width=\linewidth]{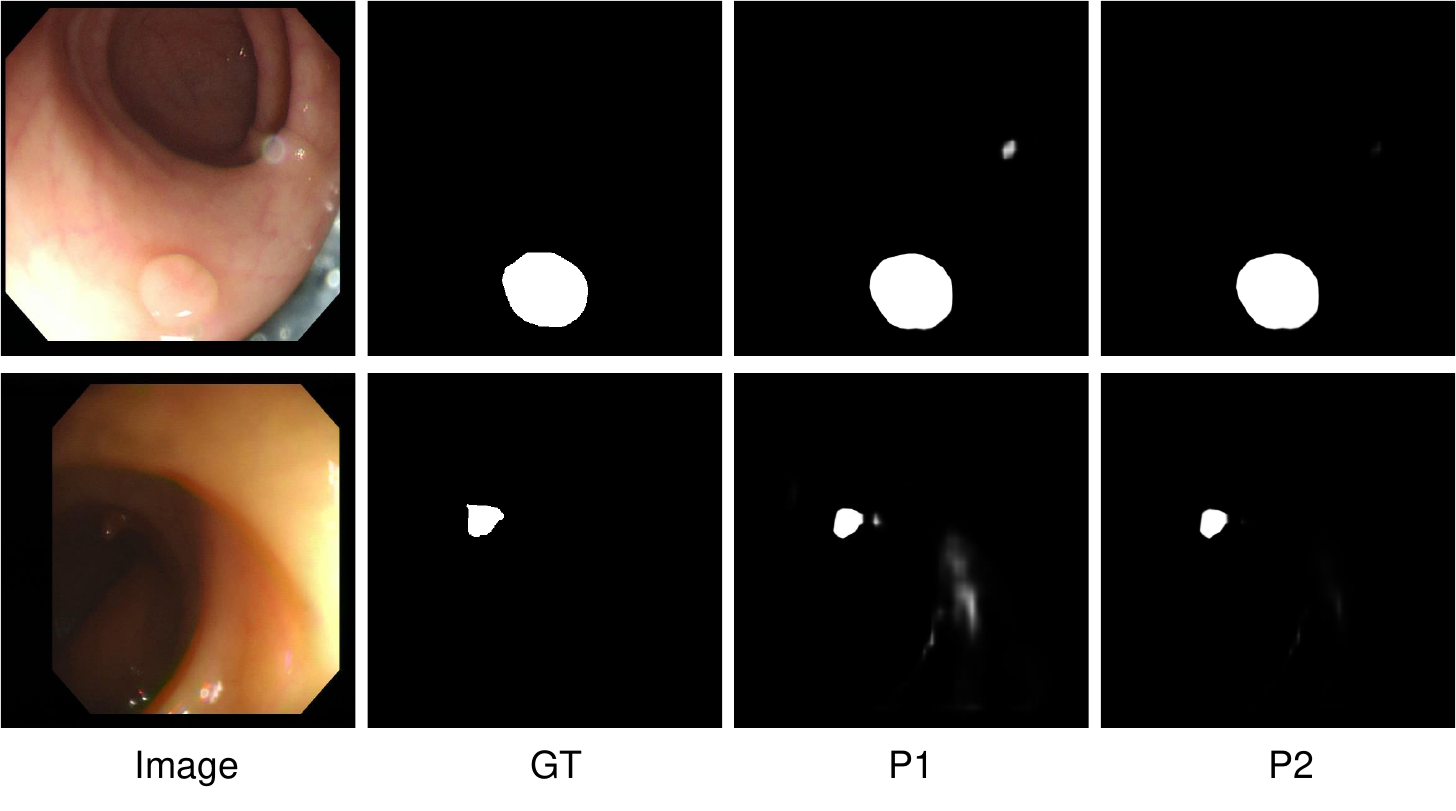}
    \end{overpic}
    \vspace{-15pt}
 	\caption{visualization of the P1 and P2 predictions.}
     \label{figure:P1P2}
     \vspace{10pt}
\end{figure}

\begin{table*}[t!]
\centering
\small
	\caption{The result of video polyp segmentation on the \emph{i.e.}, CVC-612-T and CVC-612-V.
	}
	\vspace{-5pt}
	\renewcommand{\arraystretch}{1.0}
	\setlength\tabcolsep{4.05pt}
    \begin{tabular}{l|ccccccc|ccccccc}
    \hline   
          & \multicolumn{7}{c|}{CVC-612-T~\cite{bernal2015wm}}                                  & \multicolumn{7}{c}{CVC-612-V~\cite{bernal2015wm}} \\
\cline{2-15} 
Model & mDic& mIoU & ${ F}_{\beta }^{w}$   & ${S}_{\alpha}$    & $m{E}_{\xi}$    & $max{E}_{\xi}$ & MAE  & mDic & mIoU & ${ F}_{\beta }^{w}$   & ${S}_{\alpha}$    & $m{E}_{\xi}$    & $max{E}_{\xi}$ & MAE \\
    \hline
    MICCAI'15 U-Net  & 0.711  & 0.618  & 0.694  & 0.810  & 0.836    & 0.853 & 0.058 & 0.709  & 0.597  & 0.680  & 0.826  & 0.855    & 0.872  & 0.023 \\
    TMI'19 UNet++ & 0.697  & 0.603  & 0.688  & 0.800  & 0.817    & 0.865 & 0.059 & 0.668  & 0.557  & 0.642  & 0.805  & 0.830    & 0.846 & 0.025  \\
    ISM'19 ResUNet++ & 0.616  & 0.512  & 0.604  & 0.727  & 0.758   & 0.760 & 0.084   & 0.750  & 0.646  & 0.717  & 0.829  & 0.877    & 0.879  & 0.023 \\
     MICCAI'20 ACSNet & 0.780  & 0.697  & 0.772  & 0.838  & 0.864    & 0.866  & 0.053 & 0.801  & 0.710  & 0.765  & 0.847  & 0.887   & 0.890  & 0.054  \\
    MICCAI'20 PraNet & 0.833  & 0.767  & 0.834  & 0.886  & 0.904    & \textbf{0.926} & 0.038  & 0.857  & 0.793  & 0.855  & 0.915  & 0.936   & 0.965   & 0.013 \\
    MICCAI'21 PNS-Net & 0.837  & 0.765  & 0.838  & 0.903  & 0.903    & 0.923  & 0.038 & 0.851  & 0.769  & 0.836  & 0.923  & 0.944    & 0.962  & 0.012 \\
    \hline
    \rowcolor{gray!30}
    \ourmodel~(Ours) & \textbf{0.846} & \textbf{0.776} & \textbf{0.850} & \textbf{0.895} & \textbf{0.908} & \textbf{0.926} & \textbf{0.037} & \textbf{0.882} & \textbf{0.810} & \textbf{0.874} & \textbf{0.924} & \textbf{0.963} & \textbf{0.967} & \textbf{0.012} \\
    \hline
    \end{tabular}%
  \label{tab:VPS1}
\end{table*}%

\begin{figure*}[t!]
 	\centering
 	\begin{overpic}[width=\textwidth]{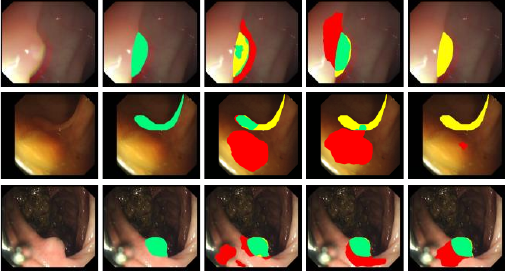}
 	\small
	\put(6,-2.5){Image}
	\put(28,-2.5){GT}
	\put(48,-2.5){\color{red}Ours}
	\put(67,-2.5){SANet}
	\put(87,-2.5){PraNet}
    \end{overpic}
    \vspace{5pt}
 	\caption{Visualization of some failure cases. Green indicates a correct polyp. Yellow is the missed polyp. Red is the wrong prediction.}
     \label{figure:limitation}
\end{figure*}

\textcolor[RGB]{31,100,212}{Effectiveness of SAM.} Similarly, we test the effectiveness of the SAM module by removing it from the overall \ourmodel~and replacing it with an element-wise addition operation, which is denoted as ``Polyp-PVT (w/o SAM)''.
The performance of the complete \ourmodel~shows an improvement of 2.9\% and 3.1\% in terms of mDic and mIoU, respectively, on ColonDB.
Fig.~\ref{figure:Visualization_ablation study} shows the benefits of SAM more intuitively.
It is found that the lack of the SAM leads to more detailed errors or even missed inspections.
As reported in Tab \ref{tab:gcn}, we add more results on the GCN in the SAM module. The experimental results further illustrate that GCN plays a key role. The effect of the lack of GCN is significantly reduced, and the effect is improved after replacing it with convolution. Still, GCN can significantly exceed the capabilities of the convolution module. The experimental results also verified the importance of GCN's large receptive field and rotation insensitivity to polyp segmentation. The rotational robustness of GCN is stronger than convolutions.  As shown in Tab \ref{tab:gcn rotation}, under the condition of large rotation (15 degrees), GCN has better adaptability to image rotation than convolutions.
To further explore the role of SAM, we visualized P1 and P2, and the results of P1 and P2 are shown in Fig~\ref{figure:P1P2}. Compared with P1, P2 has higher reliability in error recognition and identification of uncertain regions. This is mainly due to the large number of low-level details collected by CIM and mining local pixels and global semantic cues from the polyp area of SAM.

\begin{table}
 \centering
	\caption{Video polyp segmentation results on the CVC-300-TV.}
	\renewcommand{\arraystretch}{1}
	\setlength\tabcolsep{2.4pt}
	\small
    \begin{tabular}{l|ccccccc}
    \hline
     & \multicolumn{7}{c}{CVC-300-TV~\cite{bernal2012towards}} \\
\cline{2-8}    Model & mDic & mIoU & ${ F}_{\beta }^{w}$   & ${S}_{\alpha}$    & $m{E}_{\xi}$   & $max{E}_{\xi}$  & MAE\\
    \hline
    U-Net & 0.631  & 0.516  & 0.567  & 0.793  & 0.826   & 0.849   & 0.027 \\
    UNet++ & 0.638  & 0.527  & 0.581  & 0.796  & 0.831   & 0.847  & 0.024 \\
    ResUNet++ & 0.533  & 0.410  & 0.469  & 0.703  & 0.718   & 0.720  & 0.052 \\
    ACSNet & 0.732  & 0.627  & 0.703  & 0.837  & 0.871    & 0.875  & 0.016 \\
    PraNet & 0.716  & 0.624  & 0.700  & 0.833  & 0.852   & 0.904 & 0.016   \\
    PNS-Net & 0.813  & 0.710  & 0.778  & 0.909  & 0.921    & 0.942 & 0.013  \\
     \hline
     \rowcolor{gray!30}
    Ours & \textbf{0.880} & \textbf{0.802} & \textbf{0.869} & \textbf{0.915} & \textbf{0.961}  & \textbf{0.965} & \textbf{0.011}\\
    \hline
    \end{tabular}
      \label{tab:VPS2}
      \vspace{10pt}
\end{table}
\subsection{Video Polyp Segmentation}
To validate the superiority of the proposed model, we conduct experiments on the video polyp segmentation datasets.
For a fair comparison, we re-train our model with the same training datasets and use the same testing set as PNS-Net~\cite{ji2021pnsnet,ji2022video}.
We compare our model on three standard benchmarks (\emph{i.e.}, CVC-300-TV~\cite{bernal2012towards}, CVC-612-T~\cite{bernal2015wm}, and CVC-612-V~\cite{bernal2015wm}) against six cutting-edge approaches, including
U-Net~\cite{ronneberger2015unet}, UNet++~\cite{zhou2018unet++}, ResUNet++~\cite{JhaSRJLHJ19}, ACSNet~\cite{zhang2020ACSN}, PraNet~\cite{fan2020pranet}, PNS-Net~\cite{ji2021pnsnet}, in \tabref{tab:VPS1} and \tabref{tab:VPS2}.
Note that PNS-Net provides all the prediction maps of the compared methods.
As seen, our method is very competitive and far ahead of the best existing model, PNS-Net, by 3.1\% and 6.7\% on CVC-612-V and CVC-300-TV, respectively, in terms of mDice.

\subsection{Limitations}
Although the proposed \ourmodel~model surpasses existing algorithms, it still performs poorly in certain cases. 
We present some failure cases in Fig.~\ref{figure:limitation}.
As can be seen, one major limitation is the inability to detect accurate polyp boundaries with overlapping light and shadow ($1^{st}$ row).
Our model can identify the location information of polyps (green mask in $1^{st}$ row), but it regards the light and shadow part of the edge as the polyp (red mask in $1^{st}$ row). 
More deadly, our model incorrectly predicts the reflective point as a polyp (red mask in $2^{nd}$ and $3^{rd}$ rows).
We notice that the reflective points are very salient in the image. 
Therefore, we speculate that the prediction may be based on only these points.
More importantly, we believe that a simple way is to convert the input image into a gray image, which can eliminate the reflection and overlap of light and shadow to assist the model in judgment.

\section{Conclusion}
In this paper, we propose a new image polyp segmentation framework, named \ourmodel, which utilizes a pyramid vision transformer backbone as the encoder to explicitly extract more powerful and robust features.
Extensive experiments show that \ourmodel~consistently outperforms all current cutting-edge models on five challenging datasets without any pre-/post-processing.
In particular, for the unseen ColonDB dataset, the proposed model reaches a mean Dice score of above 0.8 for the first time.
Interestingly, we also surpass the current cutting-edge PNS-Net in terms of the video polyp segmentation task, demonstrating excellent learning ability. 
Specifically, we obtain the above-mention achievements by introducing three simple components, \emph{i.e.}, a cascaded fusion module (CFM), a camouflage identification module (CIM), and a similarity aggregation module (SAM), which effectively extract high and low-level cues separately, and effectively fuse them for the final output. 
We hope this research will stimulate more novel ideas for solving the polyp segmentation task.

{
\bibliographystyle{IEEEtran}
\bibliography{polyp}

\begin{thebibliography}{10}
\providecommand{\url}[1]{#1}
\csname url@samestyle\endcsname
\providecommand{\newblock}{\relax}
\providecommand{\bibinfo}[2]{#2}
\providecommand{\BIBentrySTDinterwordspacing}{\spaceskip=0pt\relax}
\providecommand{\BIBentryALTinterwordstretchfactor}{4}
\providecommand{\BIBentryALTinterwordspacing}{\spaceskip=\fontdimen2\font plus
\BIBentryALTinterwordstretchfactor\fontdimen3\font minus
  \fontdimen4\font\relax}
\providecommand{\BIBforeignlanguage}[2]{{%
\expandafter\ifx\csname l@#1\endcsname\relax
\typeout{** WARNING: IEEEtran.bst: No hyphenation pattern has been}%
\typeout{** loaded for the language `#1'. Using the pattern for}%
\typeout{** the default language instead.}%
\else
\language=\csname l@#1\endcsname
\fi
#2}}
\providecommand{\BIBdecl}{\relax}
\BIBdecl

\bibitem{fiori2014complete}
M.~Fiori, P.~Mus{\'e}, and G.~Sapiro, ``A complete system for candidate polyps
  detection in virtual colonoscopy,'' \emph{IJPRAI}, vol.~28, no.~07, p.
  1460014, 2014.

\bibitem{mamonov2014automated}
A.~V. Mamonov, I.~N. Figueiredo, P.~N. Figueiredo, and Y.-H.~R. Tsai,
  ``Automated polyp detection in colon capsule endoscopy,'' \emph{IEEE TMI},
  vol.~33, no.~7, pp. 1488--1502, 2014.

\bibitem{maghsoudi2017superpixel}
O.~H. Maghsoudi, ``Superpixel based segmentation and classification of polyps
  in wireless capsule endoscopy,'' in \emph{IEEE SPMB}, 2017.

\bibitem{ronneberger2015unet}
O.~Ronneberger, P.~Fischer, and T.~Brox, ``U-net: Convolutional networks for
  biomedical image segmentation,'' in \emph{MICCAI}, 2015.

\bibitem{fan2020pranet}
D.-P. Fan, G.-P. Ji, T.~Zhou, G.~Chen, H.~Fu, J.~Shen, and L.~Shao, ``Pranet:
  Parallel reverse attention network for polyp segmentation,'' in
  \emph{MICCAI}, 2020.

\bibitem{guo2020learn}
X.~Guo, C.~Yang, Y.~Liu, and Y.~Yuan, ``Learn to threshold: Thresholdnet with
  confidence-guided manifold mixup for polyp segmentation,'' \emph{IEEE TMI},
  vol.~40, no.~4, pp. 1134--1146, 2020.

\bibitem{wei2021shallow}
J.~Wei, Y.~Hu, R.~Zhang, Z.~Li, S.~K. Zhou, and S.~Cui, ``Shallow attention
  network for polyp segmentation,'' in \emph{MICCAI}, 2021.

\bibitem{bernal2015wm}
J.~Bernal, F.~J. S{\'a}nchez, G.~Fern{\'a}ndez-Esparrach, D.~Gil,
  C.~Rodr{\'\i}guez, and F.~Vilari{\~n}o, ``Wm-dova maps for accurate polyp
  highlighting in colonoscopy: Validation vs. saliency maps from physicians,''
  \emph{CMIG}, vol.~43, pp. 99--111, 2015.

\bibitem{silva2014toward}
J.~Silva, A.~Histace, O.~Romain, X.~Dray, and B.~Granado, ``Toward embedded
  detection of polyps in wce images for early diagnosis of colorectal cancer,''
  \emph{IJCARS}, vol.~9, no.~2, pp. 283--293, 2014.

\bibitem{tajbakhsh2015automated}
N.~Tajbakhsh, S.~R. Gurudu, and J.~Liang, ``Automated polyp detection in
  colonoscopy videos using shape and context information,'' \emph{IEEE TMI},
  vol.~35, no.~2, pp. 630--644, 2015.

\bibitem{fan2021concealed}
D.-P. Fan, G.-P. Ji, M.-M. Cheng, and L.~Shao, ``Concealed object detection,''
  \emph{IEEE TPAMI}, 2021.

\bibitem{fan2020camouflaged}
D.-P. Fan, G.-P. Ji, G.~Sun, M.-M. Cheng, J.~Shen, and L.~Shao, ``Camouflaged
  object detection,'' in \emph{CVPR}, 2020.

\bibitem{jha2020kvasir}
D.~Jha, P.~H. Smedsrud, M.~A. Riegler, P.~Halvorsen, T.~de~Lange, D.~Johansen,
  and H.~D. Johansen, ``Kvasir-seg: A segmented polyp dataset,'' in \emph{MMM},
  2020.

\bibitem{vazquez2017benchmark}
D.~V{\'a}zquez, J.~Bernal, F.~J. S{\'a}nchez, G.~Fern{\'a}ndez-Esparrach, A.~M.
  L{\'o}pez, A.~Romero, M.~Drozdzal, and A.~Courville, ``A benchmark for
  endoluminal scene segmentation of colonoscopy images,'' \emph{JHE}, vol.
  2017, 2017.

\bibitem{rahim2020survey}
T.~Rahim, M.~A. Usman, and S.~Y. Shin, ``A survey on contemporary
  computer-aided tumor, polyp, and ulcer detection methods in wireless capsule
  endoscopy imaging,'' \emph{CMIG}, p. 101767, 2020.

\bibitem{he2016deep}
K.~He, X.~Zhang, S.~Ren, and J.~Sun, ``Deep residual learning for image
  recognition,'' in \emph{CVPR}, 2016.

\bibitem{simonyan2015very}
K.~Simonyan and A.~Zisserman, ``Very deep convolutional networks for
  large-scale image recognition,'' in \emph{ICLR}, 2015.

\bibitem{li2019selective}
X.~Li, W.~Wang, X.~Hu, and J.~Yang, ``Selective kernel networks,'' in
  \emph{CVPR}, 2019.

\bibitem{wang2018mixed}
W.~Wang, X.~Li, T.~Lu, and J.~Yang, ``Mixed link networks,'' in \emph{IJCAI},
  2018.

\bibitem{long2015fully}
J.~Long, E.~Shelhamer, and T.~Darrell, ``Fully convolutional networks for
  semantic segmentation,'' in \emph{CVPR}, 2015.

\bibitem{cai2022using}
L.~Cai, M.~Wu, L.~Chen, W.~Bai, M.~Yang, S.~Lyu, and Q.~Zhao, ``Using guided
  self-attention with local information for polyp segmentation,'' in
  \emph{MICCAI}.\hskip 1em plus 0.5em minus 0.4em\relax Springer, 2022.

\bibitem{tomar2022tganet}
N.~K. Tomar, D.~Jha, U.~Bagci, and S.~Ali, ``Tganet: Text-guided attention for
  improved polyp segmentation,'' in \emph{MICCAI}.\hskip 1em plus 0.5em minus
  0.4em\relax Springer, 2022.

\bibitem{zhang2022lesion}
R.~Zhang, P.~Lai, X.~Wan, D.-J. Fan, F.~Gao, X.-J. Wu, and G.~Li,
  ``Lesion-aware dynamic kernel for polyp segmentation,'' in
  \emph{MICCAI}.\hskip 1em plus 0.5em minus 0.4em\relax Springer, 2022.

\bibitem{shi2022polyp}
J.-H. Shi, Q.~Zhang, Y.-H. Tang, and Z.-Q. Zhang, ``Polyp-mixer: An efficient
  context-aware mlp-based paradigm for polyp segmentation,'' \emph{IEEE TCSVT},
  2022.

\bibitem{zhao2022semi}
X.~Zhao, Z.~Wu, S.~Tan, D.-J. Fan, Z.~Li, X.~Wan, and G.~Li, ``Semi-supervised
  spatial temporal attention network for video polyp segmentation,'' in
  \emph{MICCAI}.\hskip 1em plus 0.5em minus 0.4em\relax Springer, 2022.

\bibitem{akbari2018polyp}
M.~Akbari, M.~Mohrekesh, E.~Nasr-Esfahani, S.~R. Soroushmehr, N.~Karimi,
  S.~Samavi, and K.~Najarian, ``Polyp segmentation in colonoscopy images using
  fully convolutional network,'' in \emph{IEEE EMBC}, 2018.

\bibitem{brandao2018towards}
P.~Brandao, O.~Zisimopoulos, E.~Mazomenos, G.~Ciuti, J.~Bernal,
  M.~Visentini-Scarzanella, A.~Menciassi, P.~Dario, A.~Koulaouzidis, A.~Arezzo
  \emph{et~al.}, ``Towards a computed-aided diagnosis system in colonoscopy:
  automatic polyp segmentation using convolution neural networks,''
  \emph{JMRR}, vol.~3, no.~02, p. 1840002, 2018.

\bibitem{zhou2018unet++}
Z.~Zhou, M.~M.~R. Siddiquee, N.~Tajbakhsh, and J.~Liang, ``Unet++: A nested
  u-net architecture for medical image segmentation,'' in \emph{DLMIA}, 2018.

\bibitem{JhaSRJLHJ19}
D.~Jha, P.~H. Smedsrud, M.~A. Riegler, D.~Johansen, T.~de~Lange, P.~Halvorsen,
  and H.~D. Johansen, ``Resunet++: An advanced architecture for medical image
  segmentation,'' in \emph{IEEE ISM}, 2019.

\bibitem{sun2019colorectal}
X.~Sun, P.~Zhang, D.~Wang, Y.~Cao, and B.~Liu, ``Colorectal polyp segmentation
  by u-net with dilation convolution,'' in \emph{IEEE ICMLA}, 2019.

\bibitem{murugesan2019psi}
B.~Murugesan, K.~Sarveswaran, S.~M. Shankaranarayana, K.~Ram, J.~Joseph, and
  M.~Sivaprakasam, ``Psi-net: Shape and boundary aware joint multi-task deep
  network for medical image segmentation,'' in \emph{IEEE EMBC}, 2019.

\bibitem{qadir2019polyp}
H.~A. Qadir, Y.~Shin, J.~Solhusvik, J.~Bergsland, L.~Aabakken, and
  I.~Balasingham, ``Polyp detection and segmentation using mask r-cnn: Does a
  deeper feature extractor cnn always perform better?'' in \emph{ISMICT}, 2019.

\bibitem{he2017mask}
K.~He, G.~Gkioxari, P.~Doll{\'a}r, and R.~Girshick, ``Mask r-cnn,'' in
  \emph{ICCV}, 2017.

\bibitem{AlamTTJR20}
S.~Alam, N.~K. Tomar, A.~Thakur, D.~Jha, and A.~Rauniyar, ``Automatic polyp
  segmentation using u-net-resnet50,'' in \emph{MediaEvalW}, 2020.

\bibitem{banik2020polyp}
D.~Banik, K.~Roy, D.~Bhattacharjee, M.~Nasipuri, and O.~Krejcar, ``Polyp-net: A
  multimodel fusion network for polyp segmentation,'' \emph{IEEE TIM}, vol.~70,
  pp. 1--12, 2020.

\bibitem{rahim2021deep}
T.~Rahim, S.~A. Hassan, and S.~Y. Shin, ``A deep convolutional neural network
  for the detection of polyps in colonoscopy images,'' \emph{BSPC}, vol.~68, p.
  102654, 2021.

\bibitem{jha2021real}
D.~Jha, S.~Ali, N.~K. Tomar, H.~D. Johansen, D.~Johansen, J.~Rittscher, M.~A.
  Riegler, and P.~Halvorsen, ``Real-time polyp detection, localization and
  segmentation in colonoscopy using deep learning,'' \emph{IEEE Access},
  vol.~9, pp. 40\,496--40\,510, 2021.

\bibitem{ahmed2020generative}
A.~M.~A. Ahmed, ``Generative adversarial networks for automatic polyp
  segmentation,'' in \emph{MediaEvalW}, 2020.

\bibitem{ThambawitaHHR20}
V.~Thambawita, S.~Hicks, P.~Halvorsen, and M.~A. Riegler,
  ``Pyramid-focus-augmentation: Medical image segmentation with step-wise
  focus,'' in \emph{MediaEvalW}, 2020.

\bibitem{tomar2020ddanet}
N.~K. Tomar, D.~Jha, S.~Ali, H.~D. Johansen, D.~Johansen, M.~A. Riegler, and
  P.~Halvorsen, ``Ddanet: Dual decoder attention network for automatic polyp
  segmentation,'' in \emph{ICPRW}, 2021.

\bibitem{huang2021hardnet}
C.-H. Huang, H.-Y. Wu, and Y.-L. Lin, ``Hardnet-mseg: A simple encoder-decoder
  polyp segmentation neural network that achieves over 0.9 mean dice and 86
  fps,'' \emph{arXiv preprint arXiv:2101.07172}, 2021.

\bibitem{chao2019hardnet}
P.~Chao, C.-Y. Kao, Y.-S. Ruan, C.-H. Huang, and Y.-L. Lin, ``Hardnet: A low
  memory traffic network,'' in \emph{CVPR}, 2019.

\bibitem{zhang2021transfuse}
Y.~Zhang, H.~Liu, and Q.~Hu, ``Transfuse: Fusing transformers and cnns for
  medical image segmentation,'' in \emph{MICCAI}, 2021.

\bibitem{yin2021duplex}
Z.~Yin, K.~Liang, Z.~Ma, and J.~Guo, ``Duplex contextual relation network for
  polyp segmentation,'' in \emph{IEEE ISBI}, 2022.

\bibitem{Zhao_2021_MICCAI}
Z.~Xiaoqi, Z.~Lihe, and L.~Huchuan, ``Automatic polyp segmentation via
  multi-scale subtraction network,'' in \emph{MICCAI}, 2021.

\bibitem{zhou2017fine}
Z.~Zhou, J.~Shin, L.~Zhang, S.~Gurudu, M.~Gotway, and J.~Liang, ``Fine-tuning
  convolutional neural networks for biomedical image analysis: actively and
  incrementally,'' in \emph{CVPR}, 2017.

\bibitem{tajbakhsh2016convolutional}
N.~Tajbakhsh, J.~Y. Shin, S.~R. Gurudu, R.~T. Hurst, C.~B. Kendall, M.~B.
  Gotway, and J.~Liang, ``Convolutional neural networks for medical image
  analysis: Full training or fine tuning?'' \emph{IEEE TMI}, vol.~35, no.~5,
  pp. 1299--1312, 2016.

\bibitem{xie2020mi}
X.~Xie, J.~Chen, Y.~Li, L.~Shen, K.~Ma, and Y.~Zheng, ``Mi\({}^{\mbox{2}}\)gan:
  Generative adversarial network for medical image domain adaptation using
  mutual information constraint,'' in \emph{MICCAI}, 2020.

\bibitem{zhang2020ACSN}
R.~Zhang, G.~Li, Z.~Li, S.~Cui, D.~Qian, and Y.~Yu, ``Adaptive context
  selection for polyp segmentation,'' in \emph{MICCAI}, 2020.

\bibitem{tomar2021automatic}
N.~K. Tomar, ``Automatic polyp segmentation using fully convolutional neural
  network,'' in \emph{MediaEvalW}, 2020.

\bibitem{JhaHEJJLRH20}
D.~Jha, S.~Hicks, K.~Emanuelsen, H.~D. Johansen, D.~Johansen, T.~de~Lange,
  M.~A. Riegler, and P.~Halvorsen, ``Medico multimedia task at mediaeval 2020:
  Automatic polyp segmentation,'' in \emph{MediaEvalW}, 2020.

\bibitem{PatelBW21}
K.~Patel, A.~M. Bur, and G.~Wang, ``Enhanced u-net: {A} feature enhancement
  network for polyp segmentation,'' in \emph{CRV}, 2021.

\bibitem{lumini2021deep}
A.~Lumini, L.~Nanni, and G.~Maguolo, ``Deep ensembles based on stochastic
  activation selection for polyp segmentation,'' in \emph{MIDL}, 2021.

\bibitem{branch2021polyp}
M.~V. Branch and A.~S. Carvalho, ``Polyp segmentation in colonoscopy images
  using u-net-mobilenetv2,'' \emph{arXiv preprint arXiv:2103.15715}, 2021.

\bibitem{khadga2021few}
R.~Khadga, D.~Jha, S.~Ali, S.~Hicks, V.~Thambawita, M.~A. Riegler, and
  P.~Halvorsen, ``Few-shot segmentation of medical images based on
  meta-learning with implicit gradients,'' \emph{arXiv preprint
  arXiv:2106.03223}, 2021.

\bibitem{sang2021ag}
D.~V. Sang, T.~Q. Chung, P.~N. Lan, D.~V. Hang, D.~Van~Long, and N.~T. Thuy,
  ``Ag-curesnest: A novel method for colon polyp segmentation,'' in \emph{IEEE
  RIVF}, 2021.

\bibitem{yang2021mutual}
C.~Yang, X.~Guo, M.~Zhu, B.~Ibragimov, and Y.~Yuan, ``Mutual-prototype
  adaptation for cross-domain polyp segmentation,'' \emph{IEEE JBHI}, 2021.

\bibitem{JhaSJLJHR21}
D.~Jha, P.~H. Smedsrud, D.~Johansen, T.~de~Lange, H.~D. Johansen, P.~Halvorsen,
  and M.~A. Riegler, ``A comprehensive study on colorectal polyp segmentation
  with resunet++, conditional random field and test-time augmentation,''
  \emph{IEEE JBHI}, vol.~25, no.~6, pp. 2029--2040, 2021.

\bibitem{jha2021nanonet}
D.~Jha, N.~K. Tomar, S.~Ali, M.~A. Riegler, H.~D. Johansen, D.~Johansen,
  T.~de~Lange, and P.~Halvorsen, ``Nanonet: Real-time polyp segmentation in
  video capsule endoscopy and colonoscopy,'' in \emph{IEEE CBMS}, 2021.

\bibitem{segtran}
S.~Li, X.~Sui, X.~Luo, X.~Xu, L.~Yong, and R.~S.~M. Goh, ``Medical image
  segmentation using squeeze-and-expansion transformers,'' in \emph{IJCAI},
  2021.

\bibitem{kim2021uacanet}
T.~Kim, H.~Lee, and D.~Kim, ``Uacanet: Uncertainty augmented context attention
  for polyp semgnetaion,'' in \emph{ACM MM}, 2021.

\bibitem{divergentNets}
V.~Thambawita, S.~A. Hicks, P.~Halvorsen, and M.~A. Riegler, ``Divergentnets:
  Medical image segmentation by network ensemble,'' in \emph{ISBI \& EndoCV},
  2021.

\bibitem{guo2021dynamic}
G.~Xiaoqing, Y.~Chen, and Y.~Yixuan, ``Dynamic-weighting hierarchical
  segmentation network for medical images,'' \emph{MIA}, p. 102196, 2021.

\bibitem{ji2021pnsnet}
G.-P. Ji, Y.-C. Chou, D.-P. Fan, G.~Chen, D.~Jha, H.~Fu, and L.~Shao,
  ``Pns-net: Progressively normalized self-attention network for video polyp
  segmentation,'' in \emph{MICCAI}, 2021.

\bibitem{vaswani2017attention}
A.~Vaswani, N.~Shazeer, N.~Parmar, J.~Uszkoreit, L.~Jones, A.~N. Gomez,
  {\L}.~Kaiser, and I.~Polosukhin, ``Attention is all you need,'' in
  \emph{NeurIPS}, 2017.

\bibitem{dosovitskiy2020image}
A.~Dosovitskiy, L.~Beyer, A.~Kolesnikov, D.~Weissenborn, X.~Zhai,
  T.~Unterthiner, M.~Dehghani, M.~Minderer, G.~Heigold, S.~Gelly \emph{et~al.},
  ``An image is worth 16x16 words: Transformers for image recognition at
  scale,'' in \emph{ICLR}, 2021.

\bibitem{pan2021scalable}
Z.~Pan, B.~Zhuang, J.~Liu, H.~He, and J.~Cai, ``Scalable visual transformers
  with hierarchical pooling,'' in \emph{ICCV}, 2021.

\bibitem{heo2021rethinking}
B.~Heo, S.~Yun, D.~Han, S.~Chun, J.~Choe, and S.~J. Oh, ``Rethinking spatial
  dimensions of vision transformers,'' in \emph{ICCV}, 2021.

\bibitem{yuan2021tokens}
L.~Yuan, Y.~Chen, T.~Wang, W.~Yu, Y.~Shi, Z.~Jiang, F.~E. Tay, J.~Feng, and
  S.~Yan, ``Tokens-to-token vit: Training vision transformers from scratch on
  imagenet,'' in \emph{ICCV}, 2021.

\bibitem{han2021transformer}
K.~Han, A.~Xiao, E.~Wu, J.~Guo, C.~Xu, and Y.~Wang, ``Transformer in
  transformer,'' \emph{Advances in Neural Information Processing Systems},
  vol.~34, pp. 15\,908--15\,919, 2021.

\bibitem{wang2021pyramid}
W.~Wang, E.~Xie, X.~Li, D.-P. Fan, K.~Song, D.~Liang, T.~Lu, P.~Luo, and
  L.~Shao, ``Pyramid vision transformer: A versatile backbone for dense
  prediction without convolutions,'' in \emph{ICCV}, 2021.

\bibitem{wang2021pvtv2}
W.~Wang, E.~Xie, X.~Li, D.-P. Fan, K.~Song, D.~Liang, T.~Lu, P.~Luo, and
  L.~Shao, ``Pvt v2: Improved baselines with pyramid vision transformer,''
  \emph{CVMJ}, vol.~8, no.~3, pp. 415--424, 2022.

\bibitem{liu2021swin}
Z.~Liu, Y.~Lin, Y.~Cao, H.~Hu, Y.~Wei, Z.~Zhang, S.~Lin, and B.~Guo, ``Swin
  transformer: Hierarchical vision transformer using shifted windows,'' in
  \emph{ICCV}, 2021.

\bibitem{cvt}
H.~Wu, B.~Xiao, N.~Codella, M.~Liu, X.~Dai, L.~Yuan, and L.~Zhang, ``Cvt:
  Introducing convolutions to vision transformers,'' in \emph{ICCV}, 2021.

\bibitem{coat}
W.~Xu, Y.~Xu, T.~Chang, and Z.~Tu, ``Co-scale conv-attentional image
  transformers,'' in \emph{ICCV}, 2021.

\bibitem{twins}
X.~Chu, Z.~Tian, Y.~Wang, B.~Zhang, H.~Ren, X.~Wei, H.~Xia, and C.~Shen,
  ``Twins: Revisiting the design of spatial attention in vision transformers,''
  \emph{Advances in Neural Information Processing Systems}, vol.~34, pp.
  9355--9366, 2021.

\bibitem{levit}
B.~Graham, A.~El-Nouby, H.~Touvron, P.~Stock, A.~Joulin, H.~J{\'e}gou, and
  M.~Douze, ``Levit: a vision transformer in convnet's clothing for faster
  inference,'' in \emph{ICCV}, 2021.

\bibitem{bhojanapalli2021understanding}
S.~Bhojanapalli, A.~Chakrabarti, D.~Glasner, D.~Li, T.~Unterthiner, and
  A.~Veit, ``Understanding robustness of transformers for image
  classification,'' in \emph{ICCV}, 2021.

\bibitem{xie2021segformer}
E.~Xie, W.~Wang, Z.~Yu, A.~Anandkumar, J.~M. Alvarez, and P.~Luo, ``Segformer:
  Simple and efficient design for semantic segmentation with transformers,''
  \emph{Advances in Neural Information Processing Systems}, vol.~34, pp.
  12\,077--12\,090, 2021.

\bibitem{Wu_2019_CVPR}
Z.~Wu, L.~Su, and Q.~Huang, ``Cascaded partial decoder for fast and accurate
  salient object detection,'' in \emph{CVPR}, 2019.

\bibitem{ioffe2015batch}
S.~Ioffe and C.~Szegedy, ``Batch normalization: Accelerating deep network
  training by reducing internal covariate shift,'' in \emph{ICML}, 2015.

\bibitem{glorot2011deep}
X.~Glorot, A.~Bordes, and Y.~Bengio, ``Deep sparse rectifier neural networks,''
  in \emph{AISTATS}, 2011.

\bibitem{woo2018cbam}
S.~Woo, J.~Park, J.-Y. Lee, and I.~So~Kweon, ``{Cbam}: Convolutional block
  attention module,'' in \emph{ECCV}, 2018.

\bibitem{hu2018squeeze}
J.~Hu, L.~Shen, and G.~Sun, ``Squeeze-and-excitation networks,'' in
  \emph{CVPR}, 2018.

\bibitem{wang2018non}
X.~Wang, R.~Girshick, A.~Gupta, and K.~He, ``Non-local neural networks,'' in
  \emph{CVPR}, 2018.

\bibitem{te2020edge}
G.~Te, Y.~Liu, W.~Hu, H.~Shi, and T.~Mei, ``Edge-aware graph representation
  learning and reasoning for face parsing,'' in \emph{ECCV}, 2020.

\bibitem{lu2019graph}
Y.~Lu, Y.~Chen, D.~Zhao, and J.~Chen, ``Graph-fcn for image semantic
  segmentation,'' in \emph{ISNN}, 2019.

\bibitem{wei2020f3net}
J.~Wei, S.~Wang, and Q.~Huang, ``F$^3$net: Fusion, feedback and focus for
  salient object detection,'' in \emph{AAAI}, 2020.

\bibitem{loshchilov2019adamw}
I.~Loshchilov and F.~Hutter, ``Decoupled weight decay regularization,'' in
  \emph{ICLR}, 2019.

\bibitem{milletari2016v}
F.~Milletari, N.~Navab, and S.-A. Ahmadi, ``V-net: Fully convolutional neural
  networks for volumetric medical image segmentation,'' in \emph{3DV}, 2016.

\bibitem{margolin2014evaluate}
R.~Margolin, L.~Zelnik-Manor, and A.~Tal, ``How to evaluate foreground maps?''
  in \emph{CVPR}, 2014.

\bibitem{Fan2021S-measure}
M.-M. Chen and D.-P. Fan, ``Structure-measure: A new way to evaluate foreground
  maps,'' \emph{IJCV}, vol. 129, pp. 2622--2638, 2021.

\bibitem{21Fan_HybridLoss}
D.-P. Fan, G.-P. Ji, X.~Qin, and M.-M. Cheng, ``Cognitive vision inspired
  object segmentation metric and loss function,'' \emph{SSI}, 2021.

\bibitem{Fan2018Emeasure}
D.-P. Fan, C.~Gong, Y.~Cao, B.~Ren, M.-M. Cheng, and A.~Borji,
  ``Enhanced-alignment measure for binary foreground map evaluation,'' in
  \emph{IJCAI}, 2018.

\bibitem{fang2019sfa}
Y.~Fang, C.~Chen, Y.~Yuan, and K.-y. Tong, ``Selective feature aggregation
  network with area-boundary constraints for polyp segmentation,'' in
  \emph{MICCAI}, 2019.

\bibitem{bernal2012towards}
J.~Bernal, J.~S{\'a}nchez, and F.~Vilarino, ``Towards automatic polyp detection
  with a polyp appearance model,'' \emph{PR}, vol.~45, no.~9, pp. 3166--3182,
  2012.

\bibitem{ji2022video}
G.-P. Ji, G.~Xiao, Y.-C. Chou, D.-P. Fan, K.~Zhao, G.~Chen, and L.~Van~Gool,
  ``Video polyp segmentation: A deep learning perspective,'' \emph{MIR},
  vol.~19, no.~06, pp. 531--549, 2022.

\end{thebibliography}
}

\end{document}